\documentclass[11pt]{article}
\usepackage{a4}
\usepackage{amsmath}
\usepackage{amssymb}
\usepackage{amsfonts}
\usepackage{graphics}
\usepackage{epsfig}
\usepackage{latexsym}
\usepackage{breqn}
\begin{document}

\noindent  arXiv:1908.07901
   \hfill  August 2019 \\

\renewcommand{\theequation}{\arabic{section}.\arabic{equation}}
\thispagestyle{empty}
\vspace*{-1,5cm}
\noindent \vskip3.3cm

\begin{center}
{\Large\bf Cubic Interaction for Higher Spins in $AdS_{d+1}$ space in the explicit covariant form}
\vspace*{1 cm}

{\large Melik Karapetyan, Ruben Manvelyan and Rubik Poghossian}

\medskip

{\small\it Yerevan Physics Institute\\ Alikhanian Br.
Str.
2, 0036 Yerevan, Armenia}\\
\medskip
{\small\tt manvel,poghos@yerphi.am; meliq.karapetyan@gmail.com}
\end{center}\vspace{2cm}

\bigskip
\begin{center}
{\sc Abstract}
\end{center}
\quad
We present a slightly modified prescription  of the radial pullback  formalism proposed previously by R. Manvelyan, R. Mkrtchyan and W. R\"uhl in 2012, where authors investigated possibility to connect the main term of higher spin interaction in flat $d+2$ dimensional space to the main term of interaction in $AdS_{d+1}$   space ignoring all trace and divergent terms but expressed directly through the $AdS$ covariant derivatives and including some curvature corrections. In this paper we succeeded to solve all necessary \emph{recurrence relations} to finalize full radial pullback of the main term of  cubic self-interaction for higher spin gauge fields in Fronsdal's formulation from flat to one dimension less  $AdS_{d+1}$ space. Nontrivial solutions of recurrence relations lead to the possibility to obtain the full set of $AdS_{d+1}$ dimensional interacting terms with all curvature corrections including trace and divergence terms from any interaction term in $d+2$ dimensional flat space.

\newpage

\section*{Introduction}
\quad
This rather technical article is devoted to cubic interaction of the higher spin gauge theory in $AdS_{d+1}$ space. So we start this introduction just pointing some interesting and important things for us sending readers for recent and not so recent reviews \cite{VasilievEqn}-\cite{Rahman:2015pzl} on the state of arts in higher spin gauge theory.

Construction of an interacting Higher Spin (HS) gauge theory  is a kind of task with some permanent background interest during more than the last thirty years starting from early work \cite{Bengtsson:1983pd}. Periodically, one can observe growing interest to this object of investigation mainly realized as some success in the construction of cubic interaction in $AdS$ or flat background and in connection with $AdS/CFT$ and  HS gravity in various dimensions. These attempts were always attractive as one more way to relate quantum theory with General Relativity and investigate HS gauge fields on the same shelf  with gravity or understand the uniqueness of gravity (spin 2 field) in comparison with other members of HS hierarchy. Because we are focused in this paper on the cubic interaction, it is worth to recall that even though consistent equations of motion \cite{VasilievEqn} for interacting higher spin fields are known for many years, the action principle for these theories remains unknown. The  usual method to construct this interacting lagrangian was to develop Fronsdal metric formalism for free fields \cite{Frons}. The crucial point here that during perturbative (Noether method) construction of interaction for HS models  we came in parallel to perturbative deformation of the free fields gauge transformation and the certain difficulties connected with the locality of the theory beyond  cubic order (see \cite{Dempster:2012vw}-\cite{Fredenhagen:2018guf} and references therein). So we see that cubic interaction up to now is the main building object of HS interaction and not all problems are solving in a fast way even on cubic level. For example, the light-cone gauge construction and classification started from the eighties of the last century for four dimensions \cite{Bengtsson:1986kh} and continued and finished by Metsaev  \cite{Metsaev} during the first decade of current century for arbitrary dimension and even with some interesting results during last years \cite{Metsaev:2019dqt} . The covariant approach went even slowly: after seminal work of Berends, Baurgers and van Dam in 1985 \cite{vanDam} and then Fradkin and Vasiliev in 1987 \cite{Fradkin} the cubic interaction and classification of vertices came to the center of interest again in 2006-2012 \cite{Bekaert:2006us}-\cite{Manvelyan:2010je}.
This development in particularly brought to interesting and elegant formulation through the generating function \cite{Sagnotti:2010at,Manvelyan:2010je} and connection with String Theory\cite{Sagnotti:2010at,Fotopoulos:2010ay}. It is worth to mention also that all these activities supplemented with the parallel development of Vasiliev's frame like formalism to cubic interaction in $AdS$ space \cite{Zinoviev}-\cite{Boulanger:2012dx}. It is interesting also in this aspects that covariant classification of cubic vertices was done for parity even dimensions $d\geq 4$ in \cite{MMR1} but classification including parity odd vertices for four and three dimensions was completed only recently in \cite{Mkrtchyan:2017ixk,Kessel:2018ugi,Conde:2016izb}.
The last point we want to mention here is that although cubic interaction in $AdS$ space has formulation developed in ambient space some years ago \cite{Joung:2011ww}-\cite{Manvelyan:2012ww} the direct formulation on the language of $AdS_{d+1}$ covariant derivatives is still unknown and realized before in \cite{Manvelyan:2012ww} for some simplest part of interaction only. From other side realization of the Noether program directly in $AdS$ space \cite{Joung:2012fv} is also extremely difficult due to noncommutativity of covariant derivatives in space with constant curvature. Therefore at the moment, the only way to see this interaction in $AdS$ space directly is to continue the approach defined in \cite{Manvelyan:2012ww}.

So the main purpose of this article to complete the first part of the program defined in \cite{Manvelyan:2012ww} where authors considered  a  version of the radial reduction (or pullback) formalism to obtain a cubic interaction of higher spin gauge fields in $AdS_{d+1}$ space from the corresponding cubic interaction in a flat $d+2$ dimensional background. The crucial point in \cite{Manvelyan:2012ww} was to write  $AdS_{d+1}$ cubic interaction terms directly in $d+1$ dimensional space using $AdS_{d+1}$ covariant derivatives. This was done there only for main term and $AdS_{d+1}$ curvature corrections without trace terms. The result was enough elegant but expressed only one simplest type of correction terms. Here we complete setup proposed in appendixes of \cite{Manvelyan:2012ww}  for all correction terms coming from main (in other words transfer and traceless) term in flat space. The key point of this paper is that we succeeded in formulation and  solution of the corresponding recurrence relations to complete radial pullback from $d+2$ dimensional flat ambient space to $AdS_{d+1}$ in all orders of curvature expansion including all possible trace terms. Another important point of this consideration is that we constructed general pullback  prescription for objects with higher derivatives of higher spin gauge fields to realize corresponding reduction for all other terms of cubic interaction pushing  this important remaining task of our program in the field of just technical work which can be done in the future without additional difficulties. This we are left for future publication.

In the first section, we presented and applied the correct radial pullback procedure for the free field reconciled with gauge invariance. Our formulation slightly differs from approaches used in  \cite{Joung:2011ww}-\cite{Manvelyan:2012ww} but completely equivalent them and more suitable for application to cubic interaction. In the second section we considered pullback for the high power of flat derivatives of HS field in $d+2$ dimensional space to power of covariant derivatives in $AdS_{d+1}$ which is the most important ingredient of cubic interaction. Doing that we solved all necessary recurrence relations arose from noncommutative algebra. In the third section using the result of previous one, we completed pullback of the main term of cubic interaction with all $AdS$ corrections supplemented by corresponding trace terms. Some technical details of calculations and useful information about cubic interaction in flat space we placed in four appendixes.

\section{Prescription for Radial Pullback  and free  HS gauge fields in $AdS$ }\setcounter{equation}{0}

\quad In this section, we present  a short review of the radial pullback  technique developed in \cite{BS, Waldron} and applied in detail to the free higher spin case in \cite{Manvelyan:2012ww}. We start  from $d+2$ dimensional flat space with coordinates  $X^{A}$ and flat $SO(1,d+1)$ invariant metric
\begin{eqnarray}
&&X^{A} \quad A=1,2,.\dots d+2 ,\label{s1.1}\\
&& ds^{2}=\eta_{AB}dX^{A}dX^{B}=-(dX^{d+2})^{2}+(dX^{d+1})^{2}+dX^{i}dX^{j}\eta_{ij} , \label{s1.2}
\end{eqnarray}
To recognize Euclidian $AdS_{d+1}$ hypersphere inside of this Ambient space we should define the following coordinate transformation
to a curvilinear coordinate system  $X^{A}\rightarrow (u,r,x^{i})$:
\begin{eqnarray}
  X^{d+2} &=& \frac{1}{2} e^{u}[r+\frac{1}{r}( L^{2}+x^{i}x^{j}\eta_{ij})] ,\nonumber\\
  X^{d+1}&=& \frac{1}{2}e^{u}[r-\frac{1}{r}(L^{2}-x^{i}x^{j}\eta_{ij})] , \nonumber\\
  X^{i} &=& e^{u}L\frac{x^{i}}{r}, \label{s1.3}\\
  - e^{2u} L^{2} &=& -(X^{d+2})^{2}+(X^{d+1})^{2}+ X^{i}X^{j}\eta_{ij} ,\label{s1.4}\\
  ds^{2}&=& L^{2}e^{2u}[-du^{2}+\frac{1}{r^{2}}(dr^{2} + dx^{i}dx^{j}\eta_{ij})] .\label{s1.5}
\end{eqnarray}
The restriction  $e^{u}=1$ leads  instead of coordinate transformations to the usual embedding of the Euclidian $AdS_{d+1}$ hypersphere with local coordinates $x^{\mu}=(x^{0},x^{i})=(r,x^{i})$ into $d+2$ dimensional flat space.

In other words, we can define the Jacobian matrix for transformation (\ref{s1.3}) in the following compact form:
\begin{eqnarray}
  E^{A}_{\mu}(u,x^{\nu}) &=&\frac{\partial X^{A}}{\partial x^{\mu}}=e^{u}e^{A}_{\mu}(x^{\nu}),\label{s1.6}\\
  E^{A}_{u}(u,x^{\nu}) &=&\frac{\partial X^{A}}{\partial u}=X^{A}(u,x^{\nu})=e^{u}L n^{A}(x^{\nu}),\label{s1.7}
\end{eqnarray}
where due to (\ref{s1.4}) the $d+1$ tangent vectors $\{e^{A}_{\mu}(x)\}^{d}_{\mu=0}$ and one normal vector $n^{A}(x)$
\begin{align}
  n^{A}(x)e^{B}_{\mu}(x)\eta_{AB} &=0\label{s1.8}\\
  n^{A}(x)n^{B}(x)\eta_{AB} &= -1\label{s1.9}
\end{align}
for embedded $AdS_{d+1}$ space define the standard induced metric $g_{\mu\nu}(x)$ and extrinsic curvature $K_{\mu\nu}(x)$ for our embedded  $AdS_{d+1}$ space:
\begin{equation}\label{s1.10}
  g_{\mu\nu}(x)=e^{A}_{\mu}(x)e^{B}_{\nu}(x)\eta_{AB}=\left(\frac{L}{x^{0}}\right)^2\delta_{\mu\nu}
\end{equation}
and
\begin{equation}\label{s1.11}
  \partial_{\mu}e^{A}_{\nu}(x)=\Gamma^{\lambda}_{\mu\nu}(g)e^{A}_{\nu}(x) + K_{\mu\nu}(x)n^{A}(x)
\end{equation}
where
\begin{eqnarray}
 \Gamma^{\lambda}_{\mu\nu}(g) &=& \Gamma^{\lambda(AdS)}_{\mu\nu}= \frac{1}{2}g^{\lambda\rho}\left(\partial_{\mu}g_{\nu\rho}+\partial_{\mu}g_{\nu\rho}-\partial_{\rho}g_{\mu\nu}\right) , \label{s1.12}\\
  K_{\mu\nu}&=& \frac{g_{\mu\nu}}{L }\label{s1.13}
\end{eqnarray}
So we see that $\Gamma^{\lambda}_{\mu\nu}(g)$ is usual Christoffel symbol constructed from induced $AdS_{d+1}$ metric and therefore we can introduce $AdS_{d+1}$ covariant derivative $\nabla_{\mu}$ and rewrite (\ref{s1.10}) in convenient form:
\begin{eqnarray}
  \nabla_{\mu}e^{A}_{\nu}(x)&=&K_{\mu\nu}(x)n^{A}(x) \label{s1.14}\\
  K_{\mu\nu}(x)&=& e^{A}_{\nu}(x)\partial_{\mu}n_{A}=- n_{A} \nabla_{\mu}e^{A}_{\nu}(x)\label{s1.15}
\end{eqnarray}

Therefor to restrict our flat theory to $AdS$ hypersphere  we should first formulate $d+2$ dimensional field theory in the curvilinear coordinates with flat
$e^{2u}(AdS_{d+1}\times \mathcal{R}_{u})$ metric
\begin{equation}\label{s1.16}
    ds^{2}=e^{2u}[-L^{2}du^{2}+ g_{\mu\nu}(x) dx^{\mu}dx^{\nu}]=G_{uu}(u)du^{2} +G_{\mu\nu}(u,x)dx^{\mu}dx^{\nu} ,
\end{equation}
where
\begin{eqnarray}
  G_{uu}(u)&=&E^{A}_{u}(u,x^{\nu})E^{B}_{u}(u,x^{\nu})\eta_{AB}=X^{A}X_{A}=-L^{2}e^{2u} \label{s1.17}\\
  G_{\mu\nu}&=&E^{A}_{\mu}(u,x^{\nu})E^{B}_{\nu}(u,x^{\nu})\eta_{AB}=e^{2u}g_{\mu\nu}(x)\label{s1.18}
\end{eqnarray}
and then define the correct prescription to go from theory in flat curvilinear space defined by Jacobian matrix $E^{A}_{u}, E^{A}_{\mu}$  to the theory with negative constant curvature on the level of $d+2 \times d+1$ embedding matrix  $e^{A}_{\mu}$ or induced metric $g_{\mu\nu}(x)$ getting rid off normal components along of $n^{A}$. The most simple check of this statement we can obtain calculating Riemann curvature of the embedded hypersphere. To perform this we should first derive differentiation rules for Frenet basis using (\ref{s1.13})-(\ref{s1.15}):
\begin{eqnarray}
  \nabla_{\mu}e^{A}_{\nu}(x)&=&\frac{g_{\mu\nu}(x)}{L }n^{A}(x)\label{s1.19}\\
  \partial_{\mu}n^{A}(x)&=& \frac{1}{L }e^{A}_{\mu}(x) ,\label{s1.20}
\end{eqnarray}
and then taking  commutator :
\begin{equation}\label{s1.21}
  [\nabla_{\mu}, \nabla_{\nu}]e^{A}_{\lambda}=R^{\quad \,\,\,\,\rho}_{\mu\nu,\lambda}e^{A}_{\rho}=K_{\lambda[\nu}K_{\mu]}^{\rho}e^{A}_{\rho}
\end{equation}
we get the standard expression for $AdS_{d+1}$ Riemann curvature and Ricci tensors
\begin{eqnarray}
  R^{\quad \,\,\,\,\rho}_{\mu\nu,\lambda} &=& -\frac{1}{L^{2}}(g_{\mu\lambda}\delta^{\rho}_{\nu}-g_{\nu\lambda}\delta^{\rho}_{\mu}) \label{s1.22}\\
   R_{\mu,\lambda} &=& -\frac{d}{L^{2}}g_{\mu\nu}, \quad R=g^{\mu\lambda}R_{\mu\lambda}=-\frac{d(d+ 1)}{L^{2}} \label{s1.23}
\end{eqnarray}
Turning to higher spins in flat ambient space  we should introduce first the following conventions.
As usual, we utilize instead of symmetric tensors such as $h^{(s)}_{A_1A_2...A_s}(X)$ polynomials homogeneous in the vector $a^{A}$ of degree $s$ at the base point $X$
\begin{equation}
h^{(s)}(X;a) = \sum_{A_{i}}(\prod_{i=1}^{s} a^{A_{i}})h^{(s)}_{A_1A_2...A_s}(X) .\label{s1.24}
\end{equation}
Then we can write the symmetrized gradient, trace, and divergence \footnote{To distinguish easily between "$a$" and "$X$" spaces we introduce the notation $\partial_{A}$ for space-time derivatives $\frac{\partial}{\partial X^{A}}$ and $\partial_{a}$ for derivatives in $a$ space.}
\begin{eqnarray}
&&Grad:h^{(s)}(X;a)\Rightarrow Gradh^{(s+1)}(X;a) = a^{A}\partial_{A}h^{(s)}(X;a) , \label{s1.25}\\
&&Tr:h^{(s)}(X;a)\Rightarrow Trh^{(s-2)}(X;a) = \frac{1}{s(s-1)}\Box_{a}h^{(s)}(X;a) ,\label{s1.26}\\
&&Div:h^{(s)}(X;a)\Rightarrow Divh^{(s-1)}(X;a) = \frac{1}{s}\eta^{AB }\partial_{A}\partial_{a^{B}}h^{(s)}(X;a) .\label{s1.27}
\end{eqnarray}
Moreover, we introduce the notation $*^{s}_a, *^{s}_b,\dots$ for a contraction in the symmetric spaces of indices $a$ or $b$
\begin{eqnarray}
  *^{s}_{a^{A}}&=&\frac{1}{(s!)^{2}} \prod^{s}_{i=1}\overleftarrow{\partial}_{a^{A_{i}}}\eta^{A_{i}B_{i}}\overrightarrow{\partial}_{a^{B_{i}}} .\label{s1.28}
\end{eqnarray}

So we should fix two important points to perform correct pullback of higher spin theory from flat ambient to one dimensional less $AdS$ space:
\begin{itemize}
      \item We should fix the ansatz for $d+2$ dimensional HS field in a way to get from one spin $s$ field exactly one spin $s$ field in $AdS_{d+1}$. The natural condition here send to zero all components normal to the embedded hypersphere
          \begin{equation}\label{s1.29}
            n^{A}h^{(s)}_{A A_2...A_s}(u,x^{\nu})\sim X^{A}(u,x^{\nu})h^{(s)}_{A A_2...A_s}(u,x^{\nu})=0
             \end{equation}
      \item Our auxiliary vector $a^{A}$ is constant in flat space
      \begin{align}
       a^{A}=&E^{A}_{u}(u,x)a^{u}(u,x^{\nu})+ E^{A}_{\mu}(u,x)a^{\mu}(u,x^{\nu})\nonumber\\=&e^{u}\left(L n^{A}(x) a^{u}(u,x)+ e^{A}_{\mu}(x)a^{\mu}(u,x)\right) \label{s1.30}\\
        \partial_{B}a^{A}&=0 ,\label{s1.31}
      \end{align}
      but in curve $AdS_{d+1}$ space there is no possibility to get covariantly constant vectors.
\end{itemize}
This means that ansatz for HS field itself is not enough for getting correct pullback for objects with  derivatives contracted with constant vector $a^{A}$. From the other side we have in hand curvilinear metric (\ref{s1.16})-(\ref{s1.18}) which we can invert and then easily invert the Jacobian matrix (\ref{s1.6})-(\ref{s1.7})
\begin{eqnarray}
 G^{uu}(u,x)&=& -\frac{e^{-2u}}{L^{2}} \label{s1.32}\\
  G^{\mu\nu}(u,x) &=& e^{-2u}g^{\mu\nu}(x)\label{s1.33} \\
 E^{u}_{A}(u,x)&=& E^{B}_{u}(u,x)\eta_{AB}G^{uu}(u,x)=-\frac{e^{-u}}{L}n_{A}(x) \label{s1.34}\\
  E^{\mu}_{A}(u,x)&=& E^{B}_{\nu}(u,x)\eta_{AB}G^{\mu\nu}(u,x)= e^{-u}e^{\mu}_{A}(x), \label{s1.35}
\end{eqnarray}
where $g^{\mu\nu}(x)$ is inverse $AdS_{d+1}$ metric and $e^{\mu}_{A}(x)=e^{B}_{\nu}(x)\eta_{AB}g^{\mu\nu}(x)$.

Then our flat-space  derivative in (\ref{s1.31}) after coordinate transformation is:
\begin{equation}\label{s1.36}
  \partial_{A}=E^{u}_{A}(u,x)\partial_{u}+E^{\mu}_{A}(u,x)\partial_{x^{\mu}}=-\frac{e^{-u}}{L}n_{A}(x)\partial_{u}+e^{-u}e^{\mu}_{A}(x) \partial_{x^{\mu}}
\end{equation}
Substituting this in (\ref{s1.31}) and taking into account (\ref{s1.30}), (\ref{s1.19}) and (\ref{s1.20})  we obtain the following four relations for derivatives of components $a^{u}(u,x), a^{\mu}(u,x)$:
\begin{eqnarray}
  &&\partial_{u}a^{u}(u,x)+a^{u}(u,x)= 0 \label{s1.37}\\
  &&\partial_{u}a^{\mu}(u,x)+a^{\mu}(u,x)= 0\label{s1.38}\\
  &&\partial_{\mu}a^{u}(u,x)+\frac{1}{L^{2}}a_{\mu}(u,x)= 0\label{s1.39}\\
   &&\nabla_{\mu}a^{\nu}(u,x)+\delta^{\nu}_{\mu}a^{u}(u,x)= 0\label{s1.40}
\end{eqnarray}
First two equations we can solve directly:
\begin{align}\label{s1.41}
  a^{u}(u,x) &= e^{-u}a^{u}(x) \\
  a^{\mu}(u,x)&= e^{-u}a^{\mu}(x)\label{s1.42}
\end{align}
Substituting these solutions in (\ref{s1.30}) and using restriction (\ref{s1.29}) we see that in curvilinear coordinates our ansatz leads to the following relation:
\begin{eqnarray}
 && h^{(s)}(X,a^{B})= h^{(s)}_{A_{1}A_{2}\dots A_{s}}(X)a^{A_{1}}a^{A_{2}}\dots a^{A_{s}}|_{X^{A}=(u, x^{\mu}), n^{A}h^{(s)}_{A\dots}=0}\nonumber\\
 && = h^{(s)}_{\mu_{1}\mu_{2}\dots \mu_{s}}(u,x)a^{\mu_{1}}(x)a^{\mu_{2}}(x)\dots a^{\mu_{s}}(x)=h^{(s)}(u,x,a^{\mu}(x))\label{s1.43}
\end{eqnarray}
where:
\begin{equation}\label{s1.44}
 h^{(s)}_{\mu_{1}\mu_{2}\dots \mu_{s}}(u,x)=h^{(s)}_{A_{1}A_{2}\dots A_{s}}(u,x)e^{A_{1}}_{\mu_{1}}(x)e^{A_{2}}_{\mu_{1}}(x)\dots e^{A_{s}}_{\mu_{s}}(x)
\end{equation}
This is correct pullback of spin s tensor field from $d+2$ dimensional flat space to $AdS_{d+1}$ space. The only reminder  about flat space we have here is $u$-dependance of $d+1$ dimensional field components in (\ref{s1.44})

The initial gauge variation of order zero in the spin $s$ field is
\begin{eqnarray}\label{s1.45}
\delta_{(0)} h^{(s)}(X^{A};a^{A})=s (a^{A}\partial_{A})\epsilon^{(s-1)}(X^{A};a^{A}) ,
\end{eqnarray}
with the traceless gauge parameter for the double traceless gauge field
\begin{eqnarray}
&&\Box_{a^{A}}\epsilon^{(s-1)}(X^{A};a^{A})=0 ,\label{s1.46}\\
&&\Box_{a^{A}}^{2} h^{(s)}(X^{A};a^{A})=0 \label{s1.47}
\end{eqnarray}
Then combining (\ref{s1.30}) and (\ref{s1.36}) we obtain due to (\ref{s1.42})
\begin{equation}\label{s1.48}
  a^{A}\partial_{A}\epsilon^{(s-1)}(X^{A};a^{A})=e^{-u}\left(a^{u}(x)\partial_{u}+a^{\mu}(x)\partial_{x^{\mu}}\right)\epsilon^{(s-1)}(u,x;a^{\mu}(x))
\end{equation}
where parameter $\epsilon^{(s-1)}(X^{A};a^{A})$ obeys to the same type  ansatz rule as the $h^{(s)}(X^{A};a^{A})$  in (\ref{s1.43})
\begin{equation}\label{s1.49}
  \epsilon^{(s-1)}(X^{A};a^{A})=\epsilon^{(s-1)}(u,x;a^{\mu}(x))
\end{equation}
The next important observation is about derivatives $\partial_{x^{\mu}} \equiv \partial_{\mu}$ in respect to $AdS_{d+1}$ coordinates $x^{\mu}$:
\begin{itemize}
  \item First note that  we mapped scalar object in flat space constructed from $X$ - dependent tensor contracted with constant vectors $a^{A}$ to the scalar object in curve space constructed from $x$-dependent tensor contracted with $x$-dependent vectors   $a^{\mu}(x)$. So as a result we obtain in r.h.s of (\ref{s1.48}) ordinary derivative $\partial_{x^{\mu}}$
  \item To see appearance of the $AdS_{d+1}$ covariant derivatives we should  use Leibnitz rule in curve space and conditions (\ref{s1.39}), (\ref{s1.40}):
      \begin{eqnarray}
       && \partial_{x^{\mu}} (T_{\nu}(x)a^{\nu}(x)) =\nabla_{\mu}T_{\nu}(x)a^{\nu}(x)+ T_{\nu}(x)\nabla_{\mu}a^{\nu}(x)\nonumber\\
       && =(\nabla_{\mu}T_{\nu}(x))a^{\nu}(x)-T_{\mu}(x)a^{u}(x)=(\nabla_{\mu}T_{\nu}(x))a^{\nu}(x)-a^{u}(x)\frac{\partial}{\partial a^{\mu}}(T_{\nu}(x)a^{\nu})\quad\quad\quad\label{s1.50}
      \end{eqnarray}
From this example we see that instead of $x$-dependent vectors we can use formally $x$-independent vectors $a^{\mu}$ (and component $a^{u}$ also ) and split $AdS$ space from formal $a^{\mu}$ space inserted only for shortening symmetric tensor contractions and symmetrizing procedures just like in the Cartesian case. But at the same time according to (\ref{s1.50}) we should replace the usual derivative with the following operators in Frenet basis:
\begin{eqnarray}
    \partial_{A} &=>& (e^{-u}\partial_{u}, e^{-u}\partial_{\mu}) ,\label{s1.51}\\
   \partial_{\mu}&=>& D_{\mu} = \nabla_{\mu}-a^{u}\partial_{a^{\mu}}-\frac{a_{\mu}}{L^{2}}\partial_{a^{u}} ,\label{s1.52}
  \end{eqnarray}
where $\nabla_{\mu}$ is $AdS$ covariant derivative constructed from the Christoffel symbols (\ref{s1.12}) with the following action rule:
\begin{equation}
    \nabla_{\mu}h^{(s)}(u,x;a)=\nabla_{\mu}h_{\mu_{1}\mu_{2}\dots \mu_{s}}(u,x)a^{\mu_{1}}a^{\mu_{2}}\dots a^{\mu_{s}}. \label{s1.53}
\end{equation}
So from now on we have instead of usual differential operator and coordinate dependent auxiliary vector components "constant" objects $a^{u}$ and $a^{\mu}$ and covariant derivative operator (\ref{s1.52}) working on rank $s$ symmetric tensors as operators working in both $x$ and $a$ spaces.
\end{itemize}
Then we can write (\ref{s1.48}) in the form:
\begin{eqnarray}\label{s1.54}
 && a^{A}\partial_{A}\epsilon^{(s-1)}(X^{A};a^{A})=e^{-u}\left(a^{u}\partial_{u}+a^{\mu}D_{\mu}\right)\epsilon^{(s-1)}(u,x;a^{\mu})\nonumber\\
 && =e^{-u}\left[a^{u}(\partial_{u}-s+1)+a^{\mu}\nabla_{\mu}\right]\epsilon^{(s-1)}(u,x;a^{\mu})
\end{eqnarray}
Using this and restricting the dependence on additional "$u$" coordinates for all fields and gauge parameters in the following (exponential) way
       \begin{eqnarray}
        &&h^{(s)}(u,x^{\mu};a^{\mu}) = e^{\Delta_{h}u}h^{(s)}(x^{\mu};a^{\mu}) ,\label{s1.55}\\
        &&\epsilon^{(s-1)}(u,x^{\mu};a^{\mu}) = e^{\Delta_{\epsilon}u}\epsilon^{(s-1)}(x^{\mu};a^{\mu}),\label{s1.56}
      \end{eqnarray}
we obtain from the (\ref{s1.45}) the following relation
\begin{eqnarray}
   &&e^{\Delta_{h}u}\delta h^{(s)}(x^{\mu};a^{\mu})= e^{(\Delta_{\epsilon}-1)u}s\left[a^{u}(\Delta_{\epsilon}-s+1)+a^{\mu}\nabla_{\mu}\right]\epsilon^{(s-1)}(x;a^{\mu}). \label{s1.57}
\end{eqnarray}
So we see that for getting from gauge transformation in $d+2$ dimensional flat space (\ref{s1.45}) the correct $AdS_{d+1}$ gauge transformation
\begin{eqnarray}
 && \delta h^{(s)}(x^{\mu};a^{\mu})= sa^{\mu}\nabla_{\mu}\epsilon^{(s-1)}(x;a^{\mu})\label{s1.58}
\end{eqnarray}
we should fix the last freedom in our ansatz in unique form
\begin{eqnarray}
  &&\Delta_{\epsilon}=s-1 \label{s1.59}\\
  && \Delta_{h}=\Delta_{\epsilon}-1 = s-2\label{s1.60}
\end{eqnarray}
which is in agreement with consideration in \cite{Joung:2011ww}-\cite{Francia:2016weg}.

After all, we can formulate our final prescription for radial pullback in the massless $AdS$ case slightly differs from our reduction formulated in   \cite{Manvelyan:2012ww}
 and can be summarized by the following three points.
  \begin{enumerate}
  \item Expand auxiliary vectors $a^{A}$ using Frenet basis for embedded $AdS$ space (\ref{s1.30}) and take into account $u$ dependents (\ref{s1.41}),(\ref{s1.42}) for components normal and tangential to the embedded hypersphere  coming from condition (\ref{s1.31}) and formal $x^{\mu}$ independence explained above. Finally, we have the following embedding rule
      \begin{equation}\label{s1.61}
       a^{A}=> L n^{A}(x) a^{u}+ e^{A}_{\mu}(x)a^{\mu}
      \end{equation}
  \item Replace  all derivatives in the following way:
      \begin{equation}\label{s1.62}
        \partial_{A}=> e^{-u}\left(-\frac{n_{A}(x)}{L}\partial_{u}+e^{\mu}_{A}(x)D_{\mu}\right)
      \end{equation}
  where $D_{\mu}$ defined in (\ref{s1.52})
      \item Restrict the dependence on additional "$u$" coordinates for all fields and gauge parameters in an exponential way with corresponding weights (\ref{s1.59}) (\ref{s1.60}) to preserve gauge invariants during pullback.
      \end{enumerate}

 Note also that our reduction rules here slightly different from rules, formulated in \cite{Manvelyan:2012ww}, especially in the area of "$u$" dependance. This happened   because we used direct solutions (\ref{s1.41}), (\ref{s1.42}) and keep derivative $\partial_{u}$ unchanged. In \cite{Manvelyan:2012ww} we removed exponential factor  $e^{-u}$ a front of derivatives and all $a^{u}$ and $a^{\mu}$vector components, replacing radial derivatives also with operator $\partial_{u}-a^{u}\partial_{a^{u}}-a^{\mu}\partial_{a^{\mu}}$ working in both $u$ and $a$ spaces. In that case scaling behaviour of field components and parameters are different from our here and in \cite{Manvelyan:2012ww} \footnote{in \cite{Manvelyan:2012ww} we had $\Delta_{h}=\Delta_{\epsilon}=2(s-1)$.}

 In any case, the final result is the same:
 After some straightforward calculation using our reduction rules we can prove that $d+2$ dimensional gauge invariant Fronsdal tensor
\begin{eqnarray}
\mathcal{F}^{(s)}(X^{A};a^{A})&&=\Box_{d+2} h^{(s)}(X^{A};a^{A})-a^{A}\partial_{A}\Big(\partial^{B}\partial_{a^{B}}h^{(s)}(X^{A};a^{A})
\nonumber\\&&-\frac{1}{2}(a^{B}\partial_{B})\Box_{a^{A}}h^{(s)}(X^{A};a^{A}\Big) , \quad\label{s1.63}
\end{eqnarray}
reduces to the $AdS_{d+1}$ gauge invariant Fronsdal tensor
\begin{eqnarray}
  &&\mathcal{F}^{(s)}(x;a^{\mu})=\Box_{d+1} h^{(s)}(x^{\mu};a^{\mu})\qquad \nonumber\\
 && -(a^{\mu}\nabla_{\mu})\Big[(\nabla^{\nu}\partial_{a^{\nu}})h^{(s)}(x;a^{\mu})
-\frac{1}{2}(a^{\nu}\nabla_{\nu})\Box_{a^{\mu}}h^{(s)}(x;a^{\mu})\Big]\quad\quad\nonumber\\
&& -\frac{1}{L^{2}}[s^{2}+s(d-5)-2(d-2)]h^{(s)}(x^{\mu};a^{\mu}))-\frac{1}{L^{2}}a^{\mu}a_{\mu}\Box_{a^{\mu}}h^{(s)}(x^{\mu};a^{\mu}).\label{s1.64}
\end{eqnarray}
in the following way
\begin{eqnarray}
 \mathcal{F}^{(s)}(X^{A};a^{A})&=& e^{(s-4)u}\mathcal{F}^{(s)}(x;a^{\mu}) ,\label{s1.65}
\end{eqnarray}
Supplementing this with the reductions for field (\ref{s1.55}), (\ref{s1.60}) and for integration volume:
\begin{eqnarray}
&&  \int d^{d+2}X =\int du d^{d+1}x \sqrt{-G}=L\int du d^{d+1}x\sqrt{g}e^{(d+2)u}\label{s1.66}
\end{eqnarray}
we obtain the following reduction rule for Fronsdal actions :
\begin{eqnarray}
  S_{0}[h^{(s)}(X^{A};a^{A})]&=&\left[L \int du e^{(d+2s-4)u}\right]\times S_{0}[h^{(s)}(x^{\mu};a^{\mu})] ,\label{s1.67}
\end{eqnarray}
where
\begin{eqnarray}
  S_{0}[h^{(s)}(X^{A};a^{A})]&=& \int d^{d+2}X\Big[ -\frac{1}{2}h^{(s)}(X^{A};a^{A})*_{a^{A}}\mathcal{F}^{(s)}(X^{A};a^{A})
    \nonumber\\&+&\frac{1}{8s(s-1)}\Box_{a^{A}}h^{(s)}(X^{A};a^{A})*_{a^{A}}\Box_{a^{A}}\mathcal{F}^{(s)}(X^{A};a^{A})\Big]  \label{s1.68}\\
   S_{0}[h^{(s)}(x^{\mu};a^{\mu})]&=&\int d^{d+1}x\sqrt{g}\Big[-\frac{1}{2}h^{(s)}(x;a^{\mu})*_{a^{\mu}}\mathcal{F}^{(s)}(x;a^{\mu})
    \nonumber\\&+&\frac{1}{8s(s-1)}\Box_{a^{\mu}}h^{(s)}(x;a^{\mu})*_{a^{\mu}}\Box_{a^{\mu}}\mathcal{F}^{(s)}(x;a^{\mu})\Big] ,\label{s1.69}
\end{eqnarray}
The overall infinite factor
\begin{equation}\label{s1.70}
    \left[L \int du e^{(d+2s-4)u}\right] ,
\end{equation}
here the same as in \cite{Manvelyan:2012ww} , where we described prescription to get correct additional $AdS$ correction terms from the full "$u$" derivative part of interaction terms. This additional terms can be found with insertion of the dimensionless delta function in measure  (\ref{s1.66}) \cite{Joung:2011ww}-\cite{Francia:2016weg}
\begin{eqnarray}
&&  \int d^{d+2}X\delta\left(\frac{\sqrt{-X^{2}}}{L}-1\right) \label{s1.71}
\end{eqnarray}
then full derivative terms will survive only for normal $u$ derivatives:
\begin{eqnarray}
&&  \int d^{d+2}X\delta\left(\frac{\sqrt{-X^{2}}}{L}-1\right)\partial^{A}\mathfrak{L}_{A}=\int d^{d+2}X\delta^{(1)}\left(\frac{\sqrt{-X^{2}}}{L}-1\right)\frac{X^{A}}{L^{2}}E^{u}_{A}\mathfrak{{L}_{u}}\nonumber\\
&&=\int du d^{d+1}x\sqrt{g}e^{(d+2)u}
\frac{\delta^{(1)}(e^{u}-1)}{L}\mathfrak{L}_{u} \label{s1.72}
\end{eqnarray}
So we see that both approaches produce the same additional corrections coming from the differentiation of overall "$u$" phase a front of full derivatives in the normal direction.
Finally, we note that this reduction procedure is more useful for investigation of interaction terms due to the very simple form of the pullback of fields and auxiliary vectors $a^{A}$ and star contractions:
\begin{eqnarray}
*^{s}_{a^{A}}&=&\frac{1}{(s!)^{2}} \prod^{s}_{i=1}\big(-\overleftarrow{\partial}_{a^{u_{i}}}\overrightarrow{\partial}_{a^{u_{i}}}
+\overleftarrow{\partial}_{a_{\mu_{i}}}\overrightarrow{\partial}_{a^{\mu_{i}}}\big)\nonumber\\
&=&\sum^{s}_{n=0}\frac{(-1)^{n}}{\binom{s}{n}}*^{n} _{a^{u}} *^{s-n} _{a^{\mu}} .\label{s1.73}
\end{eqnarray}

\section{Pullback for Power of Derivatives of HS fields from flat to embedded $AdS$ space}
\setcounter{equation}{0}
In this section, we discuss radial pullback for Cubic interaction for higher spins in a covariant off-shell formulation derived In \cite{MMR1},\cite{MMR2}. This result for flat space is in full agreement with light cone gauge results of Metsaev \cite{Metsaev}. Moreover this agreement shows that all interactions of higher spin gauge fields with any spin $s_{1}, s_{2}, s_{3}$ both in flat space and in dS or AdS are unique up to partial integration and field redefinition\footnote{This was already proven for some low spin cases of both the Fradkin-Vasiliev vertex for $2, s, s$ and the nonabelian vertex for $1, s, s$  in \cite{Boulanger:2008tg}.}
The formulation of the cubic interactions for higher spin fields in ambient space was considered in several papers \cite{Joung:2011ww}-\cite{Joung:2012fv},
In \cite{Manvelyan:2012ww} we investigated the possibility to connect the main term of interaction in flat $d+2$ dimensional space to the main term of interaction in $AdS_{d+1}$   space one dimension lower ignoring all trace and divergent terms but expressed directly through the $AdS$ covariant derivatives and including some  curvature corrections. In this article, we perform one important step forward solving task for flat main term completely and presenting full reduction or pullback including all trace and other related terms coming from  main term of cubic interaction in direct $AdS_{d+1}$ covariant form. we put in appendix A short review for the main term of cubic interaction formulated in details in \cite{MMR1,MMR2} and start here from the more convenient for radial pullback form described in \cite{Manvelyan:2012ww} where we reformulated the main term of cubic interaction (\ref{A.2}), (\ref{A.3}) in the following way
\begin{eqnarray}
&&\mathcal{L}_{I}^{main}(h^{(s_{1})}(X,a^{A}),h^{(s_{2})}(X,b^{A}),h^{(s_{3})}(X,c^{A}))
=\nonumber\\ &&\sum_{Q_{ij}} C_{Q_{12},Q_{23},Q_{31}}^{s_{1},s_{2},s_{3}}\int d^{d+2}X *^{Q_{31}+n_{3}}_{c^{A}}K^{(s_{1})}(Q_{31},n_{3};c^{A},a^{A};X)\nonumber\\
&&*^{Q_{12}+n_{1}}_{a^{A}}K^{(s_{2})}(Q_{12},n_{1};a^{A},b^{A};X)*^{Q_{23}+n_{2}}_{b^{A}}
K^{(s_{3})}(Q_{23},n_{2};b^{A},c^{A};X) ,
\quad\quad\quad\nonumber\\\label{s2.1}
\end{eqnarray}
where
\begin{eqnarray}
&&K^{(s_{1})}(Q_{12},n_{1};a^{A},b^{A};X)=(a^{A}\partial_{b^{A}})^{Q_{12}}
(a^{B}\partial_{B})^{n_{1}}h^{(s_{1})}(X;b^{C}) .\label{s2.2}
\end{eqnarray}
The most important advantage of this form that here  we can express our cubic interaction as a cube of above bitensor function
with cyclic index contraction.
From now on we put $AdS$ radius $L=1$ and use for shortness the brackets $(\dots,\dots)$ for $AdS_{d+1}$ index summation. In other words
\begin{eqnarray}
  (a,\partial_{b}) &=& a^{\mu}\partial_{b^{\mu}} ,\label{s2.3}\\
  (a,\nabla)&=&a^{\mu}\nabla_{\mu} ,\label{s2.4}
\end{eqnarray}
and
\begin{eqnarray}
  (a,D)&=&a^{\mu}D_{\mu} .\label{s2.5}
\end{eqnarray}

Another important point here is the difference in the definition of the covariant differentiation operator
(\ref{s1.52}) in the case of interaction. The minimal object here is a bitensor (\ref{s2.2}) which has two sets of symmetrized indices. In this case, we should define covariant differentiation operators for both sets of indices:
\begin{eqnarray}
    D_{\mu} &=& \nabla_{\mu}-a^{u}\partial_{a^{\mu}}-a_{\mu}\partial_{a^{u}}-b^{u}\partial_{b^{\mu}}-b_{\mu}\partial_{b^{u}} .\label{s2.6}
  \end{eqnarray}
and in a similar way for other sets of indices.
Now we have all ingredients to start analyzing the "u"- dependence  of interaction Lagrangian (\ref{s2.1}) in curvilinear coordinates (\ref{s1.3}). First of all we note that in the new frame only the measure and derivatives create additional $u$ phase (\ref{s1.66}) and (\ref{s1.62}) in addition to the three similar phase (\ref{s1.60}) coming from reduced fields. Finally, we get
\begin{eqnarray}
&&d+2+\sum^{3}_{i=1}(\Delta_{h^{(s_{i})}}-n_{i})=\sum^{3}_{i=1}(s_{i})-\Delta+d-4\label{s2.7}
\end{eqnarray}
where $\Delta$ is the number of derivatives in interaction. Then inserting minimal number of derivatives  from (\ref{A.8}) we see that our interaction rescales as\footnote{In the case of three spins ordered as $s_{1}\geq s_{2}\geq s_{3}$}
\begin{eqnarray}
   \sum^{3}_{i=1} s_{i}- \Delta_{min}+d-4 &=&d+2s_{3}-4 \label{s2.8}
\end{eqnarray}
with the obvious limit $d+2s-4$ in the self-interacting case $s_{1}=s_{2}=s_{3}=s$. So we see that the cubic interaction in the case of the minimal number of derivatives is relevant for the radial reduction procedure described in the previous section. Therefore it should produce the right curvature corrections for the main term of the cubic interaction in $AdS_{d+1}$.

\subsection*{Noncommutative algebra and $a^{u}$ stripping}
In this subsection, we consider a possible radial pullback scheme  for the main object of cubic interaction (\ref{s2.1}): the bitensorial function
\begin{eqnarray}
&&K^{(s)}(Q,n;a^{A},b^{A};X)
=(a^{A}\partial_{b^{A}})^{Q}
(a^{B}\partial_{B})^{n}h^{(s)}(X;b^{C}) .\label{s2.9}
\end{eqnarray}
This term should generate all $AdS$ curvature corrections coming from main term.
For that we study these operators in a representation that act on pullback HS field
\begin{equation}
h^{(s)}(X;b^{A})|_{X=X(u,x)} = h^{(s)}(u,x^{\mu};b^{\mu})=e^{(s-2)u}h^{(s)}(x^{\mu};b^{\mu}) .\label{s2.10}
\end{equation}
Then we can obtain these $AdS$ corrections expanding all flat $d+2$ dimensional objects in Frenet basis or in other words in term of $d+1$ dimensional $AdS$ space derivatives and vectors and normal components surviving after applying our ansatz rules:
\begin{eqnarray}
&&(a^{B}\partial_{B})^{n}|_{X=X(u,x)}=\left[e^{-u}(a^{u}\partial_{u}+a^{\mu}D_{\mu})\right]^{n}  \label{s2.11}\\
&& a^{\mu}D_{\mu}=(a,D)=(a,\nabla)-a^{u}(a,\partial_{a})-b^{u}(a,\partial_{b})-a^{2}\partial_{a^{u}}-(a,b)\partial_{b^{u}}\label{s2.12}\\
&&\textnormal{where \quad\quad} a^{2}= (a,a)=a^{\mu}a^{\nu}g_{\mu\nu}(x)\nonumber
\end{eqnarray}
and contracting over all $a^{u}, b^{u}, c^{u}$.

So we must deal with the  $d+1$ dimensional expansion for the $n$'th power of $d+2$ dimensional derivatives (\ref{s2.11}),
where the operator
\begin{eqnarray}
a^{u}\partial_{u}+a^{\mu}D_{\mu} &=&a^{\mu}\hat{\nabla}_{\mu}(g)-R ,\label{s2.13}\\
\hat{\nabla}_{\mu} &=& \nabla_{\mu} - b^{u}\partial_{b^{\mu}} - b_{\mu}\partial_{b^{u}},\label{s2.14}\\
 R &=& a^{u}[(a \partial_{a})-\partial_{u}]+a^2\partial _{a^{u}},\label{s2.15}
\end{eqnarray}
act on ground states (\ref{s2.10}). These ground states can be characterized by the total symmetry in the argument and by the fact that
they are annihilated by the following  operators:
\begin{eqnarray}
  &&\mid 0>=e^{(s-2)u}h^{(s)}(x^{\mu};b^{\mu}) \label{s2.16}\\
  &&\partial_{a^{\mu}}\mid 0> = \partial_{a^u}\mid 0> = \partial_{b^u}\mid 0> = 0 ,\label{s2.17}\\
  &&R\mid 0> = (2-s)a^{u}\mid 0> .\label{s2.18}
\end{eqnarray}
The operator of interest is
\begin{equation}
\left[e^{-u}(a,\hat{\nabla})- e^{-u} R\right]^{n},
\label{s2.19}
\end{equation}
where in the sequel it is advantageous to write the operator $R$
in the following way
\begin{eqnarray}
R&=& a^u[(a\partial_{a})+a^{u}\partial_{a^{u}}-\partial_{u}] + (a^2 - (a^u)^2)\partial_{a^u} \label{s2.20}
\end{eqnarray}
with the following important algebraic relations:
\begin{eqnarray}
&&[(a\partial_{a})+a^{u}\partial_{a^{u}},R]=R ,\label{s2.21}\\
&&[(a\partial_{a})+a^{u}\partial_{a^{u}},(a,\hat{\nabla})]=(a,\hat{\nabla}) ,\label{s2.22}\\
 &&[R,e^{-u}(a,\hat{\nabla})]=2 e^{-u}a^{u}(a,\hat{\nabla}) .\label{s2.23}
\end{eqnarray}

We have to evaluate (\ref{s2.19}) on the ground state (\ref{s2.16}). For that
Expanding this operator power (\ref{s2.19}) into a noncommutative binomial series we get
\begin{eqnarray}
&&[(a,e^{-u}\hat{\nabla})-e^{-u}R]^{n}\mid 0> = \sum_{p=0}^{n}(-1)^{p}\nonumber\\
&&\sum_{n-p \geq i_{p}\geq i_{p-1}\geq i_{p-2}...\geq i_{1} \geq 0}  (a,e^{-u}\hat{\nabla})^{n-p-i_{p}}e^{-u}R(a,e^{-u}\hat{\nabla})^{i_{p}-i_{p-1}} \dots e^{-u}R(a,e^{-u}\hat{\nabla})^{i_1}\mid 0> .\qquad\qquad\nonumber\\\label{s2.24}
\end{eqnarray}
Then using relation
\begin{equation}\label{s2.25}
    [R,(a,e^{-u}\hat{\nabla})^{i_{k}}]=2i_{k}e^{-i_{k}u} a^{u}(a,\hat{\nabla})^{i_{k}} ,
\end{equation}
we can rewrite (\ref{s2.24}) in the following form
\begin{eqnarray}
&&[(a,e^{-u}\hat{\nabla})-e^{-u}R]^{n}\mid 0> = \sum_{p=0}^{n} (-1)^{p}(a, \hat{\nabla})^{n-p}e^{(p-n)u} \nonumber\\
&&\sum_{n-p\geq i_{p}\geq i_{p-1}\geq i_{p-2}...\geq i_1\geq 0}
e^{-u}(2i_{p} a^u +R)e^{-u}(2i_{p-1}a^u +R)... e^{-u}(2i_1a^u +R)e^{(s-2)u}h^{(s)}(x^{\mu};b^{\mu}) .\nonumber\\\label{s2.26}
\end{eqnarray}
Then introducing the new objects
\begin{equation}\label{s2.27}
    \phi_{i_{k}}=2i_{k}a^{u}+R=a^{u}[2i_{k}+(a,\partial_{a})+a^{u}\partial_{a^{u}}-\partial_{u}]+[a^{2}-(a^{u})^{2}]\partial_{a^{u}} .
\end{equation}
and taking into account that
\begin{equation}\label{s2.28}
    [(a,\partial_{a})+a^{u}\partial_{a^{u}}-\partial_{u}]e^{-nu}f^{(m)}(a^{\mu},a^{u})
    =(m+n)e^{-nu}f^{(m)}(a^{\mu},a^{u}) ,
\end{equation}
we obtain
\begin{eqnarray}
&&[(a,e^{-u}\hat{\nabla})-e^{-u}R]^{n}\mid 0> = e^{(s-2-n)u}\sum_{p=0}^{n} (-1)^{p}(a, \hat{\nabla})^{n-p} \qquad\qquad\qquad \nonumber\\
&&\sum_{n-p\geq i_{p}\geq i_{p-1}\geq i_{p-2}...\geq i_1\geq 0}\phi_{i_{p}}\phi_{i_{p-1}}\dots\phi_{i_{2}}\phi_{i_{1}}h^{(s)}(x^{\mu};b^{\mu}) ,
\label{s2.29}
\end{eqnarray}
where we have $\phi_{i_{k}}$ as a very simple "creation" operators
\begin{equation}\label{s2.30}
    \phi_{i_{k}}=a^{u}[2(i_{k}+k)-s]+[a^{2}-(a^{u})^{2}]\partial_{a^{u}} .
\end{equation}

Now we show how to perform summation in (\ref{s2.29}) and obtain wanted expansion on the power of $a^{u}$ to contract after.
Introducing notation
\begin{equation}
V^{p+1}(i_{p+1})h^{(s)}(x^{\mu};b^{\mu})=\sum_{i_{p+1}\geq i_{p}\geq i_{p-1}\geq i_{p-2}...\geq i_1\geq 0}\phi_{i_{p}}\phi_{i_{p-1}}\dots\phi_{i_{2}}\phi_{i_{1}}h^{(s)}(x^{\mu};b^{\mu}) ,
\label{s2.31}
\end{equation}
and performing summation over the labels $\{i_{k}\}|^{p}_{k=1}$ we should obtain a polynomial in $a^u$ and $(a^2)$ of the form
\footnote{Note that $[p/2]$ is integer part of $p/2$ and at the end we have to insert $i_{p+1} = n-p$}
\begin{equation}
V^{p+1}(i_{p+1})=\sum_{k=0}^{[\frac{p}{2}]} \xi^{p+1}_{k}(i_{p+1}) (a^2)^{k} (a^{u}) ^{p-2k}.
\label{s2.32}
\end{equation}
Considering the last expression as an ansatz for equation
\begin{equation}\label{s2.33}
 V^{p+1}(i_{p+1})= \sum^{i_{p+1}}_{i_{p=0}}\phi_{i_{p}} V^{p}(i_{p})
\end{equation}
and using (\ref{s2.30}) we obtain the following recurrence relation for $ 2p-k$ order polynomials coefficients  $\xi^{p+1}_{k}(i_{p+1})\sim (i_{p+1})^{2p-k}+\dots$
\begin{eqnarray}
&& \xi^{p+1}_{k}(j)=\sum^{j}_{i=0}(2i+p+1+2k-s)\xi^{p}_{k}(i)+ \sum^{j}_{i=0}(p+1-2k)\xi^{p}_{k-1}(i)\label{s2.34}
\end{eqnarray}
This equation is easier to consider in "differential" form
\begin{eqnarray}
&& \xi^{p+1}_{k}(i)- \xi^{p+1}_{k}(i-1)=(2i+p+1+2k-s)\xi^{p}_{k}(i)+ (p+1-2k)\xi^{p}_{k-1}(i)  \qquad\quad\label{s2.35}
\end{eqnarray}
In Appendix B we presented solutions of latter equation obtained by direct calculation of $V^{p+1}$  using (\ref{s2.31}) for $p=1,2,3,4,\dots$.
Investigating these we arrive to the following important ansatz for $\xi^{p+1}_{k}(i)$
\begin{eqnarray}
&& \xi^{p+1}_{k}(i)=\frac{1}{(p-2k)!}(i+1)_{p}(2k+2+i-s)_{p-2k}P_{k}(i) \label{s2.36}
\end{eqnarray}
where $P_{k}(i)\sim i^{k}+\dots$ is now \emph{$p$- independent polynomial of order $k$}  and we introduced Pochhammer symbols\footnote{for falling factorial we use in this paper another notation $[s]_{n}=s(s-1)\dots(s-n+1)$}
\begin{equation}\label{s2.37}
  (a)_{n}=\frac{\Gamma(a+n)}{\Gamma(a)}=a(a+1)\dots (a+n-1)
\end{equation}
Inserting (\ref{s2.36}) in equation (\ref{s2.35}) we obtain equation for $P_{k}(i)$:
\begin{equation}\label{s2.38}
  (i+2k)P_{k}(i)-i P_{k}(i-1)=(i+2k-s)P_{k-1}(i)
\end{equation}
Then after more convenient normalization of our polynomials with additional $2k$ order factor:
\begin{equation}\label{s2.39}
\mathcal{P}_{k}(i)\equiv (i+1)_{2k}P_{k}(i)
\end{equation}
we arrive to the following simple equation with boundary condition:
\begin{eqnarray}
  &&\mathcal{P}_{k}(i)-\mathcal{P}_{k}(i-1)=(i+2k-1)(i+2k-s)\mathcal{P}_{k-1}(i) \label{s2.40}\\
  && \mathcal{P}_{0}(i)=P_{0}(i)=1 \label{s2.41}
\end{eqnarray}
This we can solve in two way: first in the form of multiple sums:
\begin{eqnarray}
&&\mathcal{P}_{k}(i)=\sum_{i\geq i_{k}\geq i_{k-1}\geq i_{k-2}...\geq i_1\geq 0}\prod^{k}_{n=1}(i_{n}+2n-1)(i_{n}+2n-s)  \label{s2.42}
\end{eqnarray}
or solving differential equation for generating function
\begin{equation}\label{s2.43}
\mathcal{P}_{k}(y)\equiv \sum^{\infty}_{i=0}\mathcal{P}_{k}(i) y^{i}
\end{equation}
where we introduced  formal variable $y$ with $|y|<1$ for production of the boundary condition:
\begin{equation}\label{s2.44}
 \mathcal{P}_{0}(y)=\sum^{\infty}_{i=0} y^{i}=\frac{1}{1-y}
\end{equation}
For this generation function, we obtain from recurrence relation (\ref{s2.40}) the equation
\begin{equation}\label{s2.45}
 (1-y)\mathcal{P}_{k}(y)=(y\frac{d}{dy}+2k-1)(y\frac{d}{dy}+2k-s)\mathcal{P}_{k-1}(y)
\end{equation}
Solving recursively and using (\ref{s2.44}) we can write the solution in the form:
\begin{equation}\label{s2.46}
  \mathcal{P}_{k}(y)=y^{-(2k+1)}\left[\frac{y^{4}}{1-y}\frac{d}{dy}y^{s}\frac{d}{dy}y^{-s}\right]^{k}\frac{y^{2}}{1-y}
\end{equation}
Finally, we can write (\ref{s2.36}) in term of $\mathcal{P}_{k}(i)$
\begin{equation}\label{s2.47}
  \xi^{p+1}_{k}(i)=\frac{1}{(p-2k)!}(2k+i+1)_{p-2k}(2k+2+i-s)_{p-2k}\mathcal{P}_{k}(i)
\end{equation}

\subsection*{Noncommutative algebra and $b^{u}$ stripping}

\quad
To extract exact dependence from $b^{u}$ and obtain final expressions written directly through the $AdS_{d+1}$ covariant derivatives $\nabla$ we have to evaluate the remaining factors
\begin{eqnarray}
&&(a, \hat{\nabla})^{n-p} = [(a,\nabla) - b^u(a,\partial_{b})-(a,b)\partial_{b^u}]^{n-p} \nonumber\\
&&= \sum_{\tilde{p}=0}^{n-p} (-1)^{\tilde{p}}{n-p \choose \tilde{p}}(a,\nabla)^{n-p-\tilde{p}}(L^{+} + L^{-})^{\tilde{p}} ,\label{s2.48}
\end{eqnarray}
where $L^{+}, L^{-}$ generate a Lie algebra
\begin{eqnarray}
&& L^{+} = b^u(a,\partial_{b}), \quad L^{-} = (a,b)\partial_{b^u} ,\label{s2.49}\\
&& [L^{+}, L^{-}] = H  = a^2 b^u \partial_{b^u} -(a,b) (a,\partial_{b}) ,\label{s2.50}\\
&& [H, L^{\pm}] = \pm 2a^2 L^{\pm} .\qquad\qquad \label{s2.51}
\end{eqnarray}
Representations of this Lie algebra are created from an ($s+1$)-dimensional vector space of "null vectors" $\{\Phi_{n}(a;b)\}|^{s}_{n=0}$
of "level" $n$
\begin{equation}
\Phi_{n}(a;b) = h^{(s)}_{\mu_1,\mu_2,...\mu_{s}} a^{\mu_1}a^{\mu_2}... a^{\mu_{n}}b^{\mu_{n+1}}b^{\mu_{n+2}}...b^{\mu_{s}},
\qquad L^{-} \Phi_{n}(a;b) = 0 ,
\label{s2.52}
\end{equation}
for any fixed tensor function $h^{s}$. From (\ref{s2.49})-(\ref{s2.51})
follows that starting from $\Phi_{0}(a;b)$ all $\Phi_{n}(a;b)$ can be produced by application of $H$
\begin{eqnarray}
&& H \Phi_0(a;b) = -s (a,b) \Phi_1(a,b) ,\label{s2.53}\\
&& H^2 \Phi_0(a;b) = [s]_2 (a,b)^2 \Phi_2(a;b)  + s a^2 (a,b) \Phi_1(a;b) ,\label{s2.54}\\
&& H^3 \Phi_0(a;b) = -\{[s]_3 (a,b)^3 \Phi_3(a;b) + 3[s]_2  a^2 (a,b)^2 \Phi_2(a;b) + s (a^2)^2 (a,b)\Phi_1(a;b)\} .\qquad\qquad\label{s2.55}
\end{eqnarray}
The ansatz
\begin{equation}
H^{n} \Phi_0(a;b) = (-1)^{n} \sum_{r=1}^{n} A_{r}^{(n)}[s]_{r}(a^2)^{n-r} (a,b)^{r}\Phi_{r}(a;b), \label{s2.56}
\end{equation}
leads to the recurrence relation
\begin{eqnarray}
&&A_{r-1}^{(n)} + r A_{r}^{(n)} = A_{r}^{(n+1)} ,\label{s2.57}\\
&&A_{r}^{(n)} = 0 \quad\textnormal{for}\quad r>n .\label{s2.58}
\end{eqnarray}
The boundary conditions $A_{-1}^{(n)}=0$ and $A_{0}^{(0)}=1$ are assumed.

Multiplying by $x^r$ and introducing
\begin{equation}
P_n(x)=\sum_{r=0}^{\infty}A_r^{(n)}x^r	\label{s2.59}
\end{equation}
we obtain simple differential equation
\begin{equation}
x \frac{d}{dx}\,\left(e^x P_n(x)\right)=e^x P_{n+1}(x)\,.\label{s2.60}
\end{equation}
which we can easily  solve since $P_0(x)=1$. Iterating $n$ times we find
\begin{equation}
e^x P_{n}(x)=\left(x \frac{d}{dx}\right)^n e^x	\,,\label{s2.61}
\end{equation}
or
\begin{equation}
P_{n}(x)=e^{-x}\left(x \frac{d}{dx}\right)^n e^x	\,.\label{s2.62}
\end{equation}
Evidently, $P_{n}(x)$ is a polynomial of order  $n$, which means that
$A_r^{(n)}=0$ for $r>n$.

Finally,  we can find a "double" generating function.
Introducing
\begin{equation}
Q(x,t)=\sum_{n=0}^{\infty}P_{n}(x)\,\frac{t^n}{n!}\label{s2.63}
\end{equation}
we see that
\begin{equation}
Q(x,t)=e^{-x}e^{tx\frac{d}{dx}} e^x=e^{x(e^t-1)}\label{s2.64}	
\end{equation}
where we have explored the fact that the operator
$
e^{t x\frac{d}{dx}}
$
rescales the variable $x$ by the factor $e^t$.

It is not difficult to get a simple combinatorial formula for $A_r^{(n)}$.
Let us denote by ${\cal P}(n,r)$ the set of partitions of $n$ into
$r$ nonzero parts. The partitions are in one to one correspondence
with Young diagrams with $n$ boxes and $r$ rows. An arbitrary partition
$\lambda$ may be represented as $\lambda =1^{k_1}2^{k_2}3^{k_2}\cdots $,
where the nonnegative integer $k_i$ indicates the number of rows with
length $i$. For example the partition $8=1+1+3+3$ is represented as $1^22^03^2$,
hence $\{k_1,k_2,k_3\}=\{2,0,2\}$ and $k_4=k_5=\cdots=0$. The corresponding
Young diagram consists of two rows of length $3$ and two rows of length $1$.
For a diagram $\lambda\in {\cal P}(n,r)$ let us arbitrarily distribute the integers
$1,2,\cdots n$ among boxes. Let us identify two configurations which differ
from each other by permutations of numbers along rows or by permutation of entire rows
of same lengths. Evidently, the number of non-equivalent distributions is given by
\begin{eqnarray}
S(\lambda)=\frac{n!}{\prod_{i\ge 1}k_i!(i!)^{k_i}}
\label{sym_factor}
\end{eqnarray}
Expanding (\ref{s2.64}) in $x$ and $t$ we get
\begin{eqnarray}
e^{x(e^t-1)}=\prod_{i=1}^\infty\sum_{k_i=0}^\infty\frac{x^{k_i}t^{ik_i}}{k_i!(i!)^{k_i}}
\label{Q_expansion}
\end{eqnarray}
Now comparing (\ref{Q_expansion}) with (\ref{sym_factor}) one easily gets
\begin{eqnarray}
A_n^{(r)}=\sum_{\lambda\in {\cal P}(n,r)}S(\lambda)
\end{eqnarray}

With the help of the basis $\{\Phi_{n}(a;b)\}_{n=0}^{s}$ of null vectors the representation of the Lie algebra can be constructed as follows.
We start from
\begin{eqnarray}
&&(L^{+} + L^{-})^{\tilde{p}} \Phi_0(b) = \sum_{\tilde{k}=0}^{\tilde{p}}\sum_{\tilde{p}-\tilde{k} \geq i_{\tilde{k}}\geq i_{\tilde{k}-1}\geq i_{\tilde{k}-2}...\geq i_{1} \geq 1}  \nonumber\\ &&(L^{+})^{\tilde{p}-\tilde{k}-i_{\tilde{k}}} L^{-}(L^{+})^{i_{\tilde{k}}-i_{\tilde{k}-1}}L^{-} (L^{+})^{i_{\tilde{k}-1} -i_{\tilde{k}-2}} L^{-}...(L^{+})^{i_2-i_1}L^{-}(L^{+})^{i_1}\Phi_0(b) \label{s2.65}
.\label{4.41}
\end{eqnarray}
Only commutators of $L^{-}$ with powers of $L^{+}$ arise
\begin{eqnarray}
&&[L^{-}, (L^{+})^{i}] = -\sum_{j = 0}^{i-1} (L^{+})^{i-j-1}H (L^{+})^j = \qquad \nonumber\\
&&-\sum_{j=0}^{i-1}(L^{+})^{i-1}( H +2j a^2)
=  - (L^{+})^{i-1}( i H + [i]_2\, a^2).
\label{s2.66}
\end{eqnarray}
Here we recognize that the whole basis $\{\Phi_{n}(a;b)\}$ of null vectors is produced from $\Phi_0(b)$ by the action of $H$. With the shorthand
\begin{equation}
\psi_{i} = i H + [i]_2\, a^2 ,
\label{s2.67}
\end{equation}
the result is
\begin{equation}
\sum_{\tilde{k}=1}^{[\frac{\tilde{p}}{2}]}(-1)^{\tilde{k}}(L^{+})^{\tilde{p}-2\tilde{k}}W^{\tilde{k}}(a^{2},H)\Phi_{0}(b)=
\sum_{\tilde{k}=1}^{[\frac{\tilde{p}}{2}]}(b^{u})^{\tilde{p}-2\tilde{k}}(-1)^{\tilde{k}}(a,\partial_{b})^{\tilde{p}-2\tilde{k}}W^{\tilde{k}}(a^{2},H)\Phi_{0}(b)
\label{s2.68}
\end{equation}
where
\begin{eqnarray}
  &&W^{\tilde{k}}(a^{2},H,i_{\tilde{k}+1})\Phi_{0}(b)= \sum_{i_{\tilde{k}+1}\geq i_{\tilde{k}}\geq i_{\tilde{k}-1}\geq i_{\tilde{k}-2}...\geq i_2 \geq i_1\geq 1} \psi_{i_{\tilde{k}}-\tilde{k}+1}\psi_{i_{\tilde{k}-1}-\tilde{k}+2}
\psi_{i_{\tilde{k}-2}-\tilde{k}+3}...\psi_{i_2-1} \psi_{i_1} \Phi_0(b). \nonumber\\\label{s2.69}
\end{eqnarray}
The sum is a homogeneous polynomial of $H$ and $a^2$ of degree $\tilde{k}$, \footnote{Remember that $H$ is second order in $a$ as well.}:
\begin{eqnarray}
  && W^{\tilde{k}}(a^{2},H,i_{\tilde{k}+1})=\sum^{\tilde{k}}_{m=0}\eta^{m}_{\tilde{k}}(i_{\tilde{k}+1})(a^{2})^{m}H^{\tilde{k}-m}\label{s2.70}
\end{eqnarray}
Using this ansatz and doing in the way similar to (\ref{s2.32}) we derive from
\begin{eqnarray}
  &&W^{\tilde{k}+1}(a^{2},H,i_{\tilde{k}+2})=\sum^{i_{\tilde{k}+2}}_{i_{\tilde{k}+1}=1}\psi_{i_{\tilde{k}+1}-\tilde{k}}W^{\tilde{k}}
  (a^{2},H,i_{\tilde{k}+1})\label{s2.71}
\end{eqnarray}
the following recurrence relation
\begin{eqnarray}
  && \eta^{m}_{\tilde{k}+1}(j)=\sum^{j}_{i=1}\left[(i-\tilde{k})\eta^{m}_{\tilde{k}}(i)+(i-\tilde{k})(i-\tilde{k}-1)\eta^{m-1}_{\tilde{k}}(i)\right]\label{s2.72}
\end{eqnarray}
or without summation:
\begin{eqnarray}
  && \eta^{m}_{\tilde{k}+1}(i)- \eta^{m}_{\tilde{k}+1}(i-1)=(i-\tilde{k})\eta^{m}_{\tilde{k}}(i)+(i-\tilde{k})(i-\tilde{k}-1)\eta^{m-1}_{\tilde{k}}(i)\label{s2.73}
\end{eqnarray}
Investigating the structure of this polynomial coefficients (See Appendix C) we can factorize again $i^{2\tilde{k}}$ terms and write in this form
\begin{equation}\label{s2.74}
 \eta^{m}_{\tilde{k}}(i) = \frac{2^{m - \tilde{k}}3^{-m}}{(\tilde{k} - m)!m!} (i-\tilde{k}+1)_{2\tilde{k}}P_m(i,\tilde{k}), \quad P_0(i,\tilde{k})=1
\end{equation}
where the polynomials $P_m(i,\tilde{k})\sim (i-\frac{\tilde{k}}{2})^{m}+\dots$ is  $m$th orders in $i$ and $\tilde{k}$ with binomial leading term and satisfy the equation
\begin{equation}
(i + \tilde{k} + 1)P_m(i,\tilde{k}+1) - (i - \tilde{k} - 1)P_m( i - 1, \tilde{k} + 1) =
2 (\tilde{k} - m + 1)P_m(i, \tilde{k}) + 3m (i -\tilde{k} - 1)P_{m - 1}( i, \tilde{k})\label{s2.75}
\end{equation}
with the same level of difficulty to solve as (\ref{s2.73}).
From the other hand  representation (\ref{s2.68}) extract $b^{u}$ dependence and we can calculate coefficients $\eta^{m}_{\tilde{k}}(i_{\tilde{k}+1})$ from (\ref{s2.69}) directly. comparing (\ref{s2.70}) with (\ref{s2.69}) and taking into account (\ref{s2.67}) we see that it is possible to write
\begin{equation}\label{s2.76}
  \eta^{m}_{\tilde{k}}(\tilde{p}-\tilde{k})=\eta^{m}_{\tilde{k}}(i_{\tilde{k}+1})|_{i_{\tilde{k}+1}=\tilde{p}-\tilde{k}}
\end{equation}
in the following form:
\begin{align}
  \eta^{m}_{\tilde{k}}(\tilde{p}-\tilde{k})&=\sum_{\tilde{p}-\tilde{k}\geq i_{\tilde{k}}\geq i_{\tilde{k}-1}\geq i_{\tilde{k}-2}...\geq i_2 \geq i_1\geq 1}\quad\quad\sum_{\tilde{k}\geq n_{m}\geq n_{m-1}\geq n_{m-2}...\geq n_2 \geq n_1\geq 1}\nonumber\\& \prod_{l_{m}= n_{m}+1}^{\tilde{k}}(i_{l_{m}}-l_{k}+1)[i_{n_{m}}-n_{m}+1]_{2}\prod_{l_{m-1}= n_{m-1}+1}^{n_{m}-1}(i_{l_{m-1}}-l_{m-1}+1)[i_{n_{m-1}}-n_{m-1}+1]_{2}
  \dots\nonumber\\
  \dots&\prod_{l_{2}= n_{2}+1}^{n_{3}-1}(i_{l_{2}}-l_{2}+1)[i_{n_{2}}-n_{2}+1]_{2}\prod_{l_{1}= n_{1}+1}^{n_{2}-1}(i_{l_{1}}-l_{1}+1)[i_{n_{1}}-n_{1}+1]_{2}\prod_{l=1}^{n_{1}-1}
  (i_{l}-l+1) \label{s2.77}
\end{align}
This formula means that we should inside of expression for $ \eta^{0}_{\tilde{k}}(\tilde{p}-\tilde{k})$:
\begin{equation}\label{s2.78}
 \eta^{0}_{\tilde{k}}(\tilde{p}-\tilde{k})= \sum_{\tilde{p}-\tilde{k}\geq i_{\tilde{k}}\geq i_{\tilde{k}-1}\geq i_{\tilde{k}-2}...\geq i_2 \geq i_1\geq 1}
  \prod_{l=1}^{\tilde{k}}
  (i_{l}-l+1)
\end{equation}
replace $m$ brackets $(i_{n_{r}}-n_{r}+1)|^{m}_{r=1}$  with the  $m$ Pochhammers $\{[i_{n_{r}}-n_{r}+1]_{2}\}|^{m}_{r=1}$ in all possible ways and then take sums.

\section{Pullback of the main term of cubic self-interaction}
\setcounter{equation}{0}
Now we start to collect things together and present all terms of cubic interaction produced from the main term in one dimension more flat space. First, we look at the main term in the case of a cubic self-interaction. This can be obtained from the general expressions (\ref{A.4})-(\ref{A.6}) taking
\begin{eqnarray}
  s_{1} &=& s_{2}=s_{3}=s ,\label{s3.1}\\
  \nu_{1} &=& \nu_{2}=\nu_{3}=0 ,\label{s3.2}\\
   Q_{23} &=& n_{1}=\alpha ,\label{s3.3}\\
   Q_{31} &=& n_{2}=\beta ,\label{s3.4}\\
   Q_{12} &=& n_{3}=\gamma .\label{s3.5}
\end{eqnarray}
Then (\ref{s2.1}), (\ref{s2.2}) transform to the following nice cyclic (in (a)$\alpha,(b)\beta,(c)\gamma$) expression with trinomial coefficients :
\begin{eqnarray}
\mathcal{L}_{I}^{main}&=&\sum_{\alpha, \beta, \gamma \atop \alpha+\beta+\gamma=s} {s \choose \alpha,\beta,\gamma} \int d^{d+2}X\nonumber\\
&&*_{a}^{\gamma+\alpha}(a^{A}\partial_{b^{A}})^{\gamma}(a^{B}\partial_{B})^{\alpha}h^{(s)}(X;b^{C})\nonumber\\
&&*_{b}^{\alpha+\beta}(b^{D}\partial_{c^{D}})^{\alpha}
(b^{E}\partial_{E})^{\beta}h^{(s)}(X;c^{F})\nonumber \\
&&*_{c}^{\beta+\gamma}(c^{G}\partial_{a^{G}})^{\beta}
(c^{H}\partial_{H})^{\gamma}h^{(s)}(X;a^{K})
,\label{s3.6}
\end{eqnarray}

The main result of the previous section is that we can expand each line of (\ref{s3.6}) and extract $a^{u}, b^{u}, c^{u}$ dependence to contract with expansion of star product and write exact expression in the term of $AdS_{d+1}$ dimensional covariant derivatives and curvature corrections. Combining (\ref{s2.29})-(\ref{s2.32}) and (\ref{s2.48})-(\ref{s2.70}) we can write\footnote{For shortening notation we introduce instead of $H(a,b)$ from (\ref{s2.50}) $H_{1}$ and then $H_{2}=H(b,c)$ and $H_{3}=H(c,a)$ correspondingly.}
\begin{eqnarray}
  && (a^{B}\partial_{B})^{\alpha}h^{(s)}(X;b^{C})=
 e^{(s-2-\alpha)u}\sum_{p_{1}=0}^{\alpha}\sum_{k_{1}=0}^{[\frac{p_{1}}{2}]}\sum_{\tilde{p_{1}}=0}^{\alpha-p_{1}}
 \sum_{\tilde{k_{1}}=1}^{[\frac{\tilde{p}_{1}}{2}]} (-1)^{p_{1}+\tilde{p_{1}}+\tilde{k_{1}}}(a^{u}) ^{p_{1}-2k_{1}} (b^{u})^{\tilde{p_{1}}-2\tilde{k_{1}}}(a,\nabla)^{\alpha-p_{1}-\tilde{p_{1}}}\nonumber\\&&\xi^{p_{1}+1}_{k_{1}}(\alpha-p_{1}){\alpha-p_{1} \choose \tilde{p_{1}}} (a^2)^{k_{1}}(a,\partial_{b})^{\tilde{p_{1}}-2\tilde{k_{1}}}W^{\tilde{k_{1}}}(a^{2},H_{1})
h^{(s)}(x^{\mu};b^{\mu})
 \label{s3.7}
\end{eqnarray}
Then expanding :
\begin{eqnarray}
  && (a^{A}\partial_{b^{A}})^{\gamma}=\sum^{\gamma}_{m=0}
  \binom{\gamma}{m_{1}}(a^{u}\partial_{b^{u}})^{m_{1}}(a,\partial_{b})^{\gamma-m_{1 }}  \label{s3.8}
\end{eqnarray}
we obtain
\begin{eqnarray}
  &&(a^{A}\partial_{b^{A}})^{\gamma}(a^{B}\partial_{B})^{\alpha}h^{(s)}(X;b^{C})=e^{(s-2-\alpha)u}
  \sum^{\gamma}_{m_{1}=0}\sum^{\gamma,\alpha,[\frac{p_{1}}{2}],\alpha-p_{1},[\frac{\tilde{p}_{1}}{2}]}_{m_{1},p_{1},k_{1},\tilde{p_{1}},\tilde{k_{1}}} \nonumber\\&&(a^{u}) ^{p_{1}-2k_{1}+m_{1}} (b^{u})^{\tilde{p_{1}}-2\tilde{k_{1}}-m_{1}}(a,\partial_{b})^{\gamma+\tilde{p_{1}}-2\tilde{k_{1}}-m_{1}}
  (a,\nabla)^{\alpha-p_{1}-\tilde{p_{1}}}\Theta[\gamma,\alpha,m_{1},p_{1},k_{1},\tilde{p_{1}},\tilde{k_{1}},a^{2},H_{1}]h^{(s)}(b^{\mu}) . \quad\quad\quad \label{s3.9}
\end{eqnarray}
where:
\begin{eqnarray}
  &&  \sum^{\gamma,\alpha,[\frac{p_{1}}{2}],\alpha-p_{1},[\frac{\tilde{p}_{1}}{2}]}_{m_{1},p_{1},k_{1},\tilde{p_{1}},\tilde{k_{1}}}= \sum^{\gamma}_{m_{1}=0}\sum_{p_{1}=0}^{\alpha}\sum_{k_{1}=0}^{[\frac{p_{1}}{2}]}
  \sum_{\tilde{p_{1}}=0}^{\alpha-p_{1}}\sum_{\tilde{k_{1}}=1}^{[\frac{\tilde{p}_{1}}{2}]}  \label{s3.10}
\end{eqnarray}
and
\begin{eqnarray}
  && \Theta[\gamma,\alpha,m_{1},p_{1},k_{1},\tilde{p_{1}},\tilde{k_{1}},a^{2},H_{1}]\nonumber\\
  &&=(-1)^{p_{1}+\tilde{p_{1}}+\tilde{k_{1}}}[\tilde{p_{1}}-2\tilde{k_{1}}]_{m_{1}}\binom{\gamma}{m_{1}}\xi^{p_{1}+1}_{k_{1}}(\alpha-p_{1}){\alpha-p_{1} \choose \tilde{p_{1}}} (a^2)^{k_{1}}W^{\tilde{k_{1}}}(a^{2},H_{1})\nonumber\\\label{s3.11}
\end{eqnarray}
Then we can write expression for the whole main interaction term
\begin{eqnarray}
  &&\mathcal{L}_{I}^{main}=\int du e^{(d+2s-4)u} d^{d+1}x\sqrt{g}\sum_{\alpha, \beta, \gamma \atop \alpha+\beta+\gamma=s} {s \choose \alpha,\beta,\gamma}
\sum^{\gamma,\alpha,[\frac{p_{1}}{2}],\alpha-p_{1},[\frac{\tilde{p_{1}}}{2}]}_{m_{1},p_{1},k_{1},\tilde{p_{1}},\tilde{k_{1}}}
\,\,\sum^{\alpha,\beta, [\frac{p_{2}}{2}],\beta-p_{2},[\frac{\tilde{p_{2}}}{2}]}_{m_{2},p_{2},k_{2},\tilde{p_{2}},\tilde{k_{2}}}
\,\,\sum^{\beta,\gamma, [\frac{p_{3}}{2}],\gamma-p_{3},[\frac{\tilde{p_{3}}}{2}]}_{m_{3},p_{3},k_{3},\tilde{p_{3}},\tilde{k_{3}}}\nonumber\\
&&\sum^{\gamma+\alpha, \alpha+\beta,\beta+\gamma}_{n_{1},n_{2},n_{3}=0}\frac{(-1)^{n_{1}+n_{2}+n_{3}}}{\binom{\gamma+\alpha}{n_{1}}\binom{\alpha+\beta}{n_{2}}
\binom{\beta+\gamma}{n_{3}}}
*^{n_{1}} _{a^{u}}*^{n_{2}} _{b^{u}}*^{n_{3}} _{c^{u}} *^{\gamma+\alpha-n_{1}} _{a^{\mu}}*^{\alpha+\beta-n_{2}} _{b^{\mu}}*^{\beta+\gamma-n_{3}} _{c^{\mu}} \nonumber\\
 && (a^{u}) ^{p_{1}-2k_{1}+m_{1}} (b^{u})^{\tilde{p_{1}}-2\tilde{k_{1}}-m_{1}}(a,\partial_{b})^{\gamma+\tilde{p_{1}}-2\tilde{k_{1}}-m_{1}}
  (a,\nabla)^{\alpha-p_{1}-\tilde{p_{1}}}\Theta[\gamma,\alpha,m_{1},p_{1},k_{1},\tilde{p_{1}},\tilde{k_{1}},a^{2},H_{1}]h^{(s)}(b^{\mu}) \nonumber\\
&&(b^{u}) ^{p_{2}-2k_{2}+m_{2}} (c^{u})^{\tilde{p_{2}}-2\tilde{k_{2}}-m_{2}}(b,\partial_{c})^{\alpha+\tilde{p_{2}}-2\tilde{k_{2}}-m_{2}}
  (b,\nabla)^{\beta-p_{2}-\tilde{p_{2}}}\Theta[\alpha,\beta, m_{2},p_{2},k_{2},\tilde{p_{2}},\tilde{k_{2}},b^{2},H_{2}]h^{(s)}(c^{\mu}) \nonumber\\
  &&(c^{u}) ^{p_{3}-2k_{3}+m_{3}} (a^{u})^{\tilde{p_{3}}-2\tilde{k_{3}}-m_{3}}(c,\partial_{a})^{\beta+\tilde{p_{3}}-2\tilde{k_{3}}-m_{3}}
  (c,\nabla)^{\gamma-p_{3}-\tilde{p_{3}}}\Theta[\beta,\gamma,m_{3},p_{3},k_{3},\tilde{p_{3}},\tilde{k_{3}},c^{2},H_{3}]h^{(s)}(a^{\mu}) \nonumber\\\label{s3.12}
\end{eqnarray}

Now we can contract all non $AdS_{d+1}$ components $a^{u}, b^{u}, c^{u}$ using corresponding $"u"$-stars  from second line of (\ref{s3.12}). This leads to the following constraints for summation indices:
\begin{eqnarray}
  &&p_{1}-2k_{1}+m_{1}=\tilde{p_{3}}-2\tilde{k_{3}}-m_{3}=n_{1} \label{s3.13}\\
  &&p_{2}-2k_{2}+m_{2}=\tilde{p_{1}}-2\tilde{k_{1}}-m_{1}=n_{2} \label{s3.14}\\
  &&p_{3}-2k_{3}+m_{3}=\tilde{p_{2}}-2\tilde{k_{2}}-m_{2}=n_{3} \label{s3.15}
\end{eqnarray}
So we can take summation over $m_{i}, i=1,2,3$ with remaining constraints on other variables :
\begin{eqnarray}
  && p_{1}+\tilde{p_{1}}=n_{1}+n_{2}+2(k_{1}+\tilde{k_{1}}) \label{s3.16}\\
  && p_{2}+\tilde{p_{2}}=n_{2}+n_{3}+2(k_{2}+\tilde{k_{2}}) \label{s3.17}\\
  && p_{3}+\tilde{p_{3}}=n_{3}+n_{1}+2(k_{3}+\tilde{k_{3}}) \label{s3.18}
\end{eqnarray}
Relations (\ref{s3.13})-(\ref{s3.15}) restrict also summation ranges for $n_{1}, n_{2}, n_{3}$ from zero to $\alpha, \beta, \gamma$.
Then we have
\begin{eqnarray}
  &&\mathcal{L}_{I}^{main}=\int du e^{(d+2s-4)u} d^{d+1}x\sqrt{g}\sum_{\alpha, \beta, \gamma \atop \alpha+\beta+\gamma=s} {s \choose \alpha,\beta,\gamma}
\sum^{\alpha,[\frac{p_{1}}{2}],\alpha-p_{1},[\frac{\tilde{p_{1}}}{2}]}_{p_{1},k_{1},\tilde{p_{1}},\tilde{k_{1}}}
\,\,\sum^{\beta, [\frac{p_{2}}{2}],\beta-p_{2},[\frac{\tilde{p_{2}}}{2}]}_{p_{2},k_{2},\tilde{p_{2}},\tilde{k_{2}}}
\,\,\sum^{\gamma, [\frac{p_{3}}{2}],\gamma-p_{3},[\frac{\tilde{p_{3}}}{2}]}_{p_{3},k_{3},\tilde{p_{3}},\tilde{k_{3}}}\nonumber\\
&&\sum^{\alpha, \beta,\gamma}_{n_{1},n_{2},n_{3}=0}\frac{(-1)^{n_{1}+n_{2}+n_{3}}}{\binom{\gamma+\alpha}{n_{1}}\binom{\alpha+\beta}{n_{2}}
\binom{\beta+\gamma}{n_{3}}}
 *^{\gamma+\alpha-n_{1}} _{a^{\mu}}*^{\alpha+\beta-n_{2}} _{b^{\mu}}*^{\beta+\gamma-n_{3}} _{c^{\mu}} \nonumber\\
 && (a,\partial_{b})^{\gamma+n_{2}}
  (a,\nabla)^{\alpha-n_{1}-n_{2}-2(k_{1}+\tilde{k_{1}})}\tilde{\Theta}[\gamma,\alpha,n_{2},p_{1},k_{1},\tilde{p_{1}},\tilde{k_{1}},a^{2},H_{1}]h^{(s)}(b^{\mu}) \nonumber\\
&&(b,\partial_{c})^{\alpha+n_{3}}
  (b,\nabla)^{\beta-n_{2}-n_{3}-2(k_{2}+\tilde{k_{2}})}\tilde{\Theta}[\alpha,\beta,n_{3},p_{2},k_{2},\tilde{p_{2}},\tilde{k_{2}},b^{2},H_{2}]h^{(s)}(c^{\mu}) \nonumber\\
  &&(c,\partial_{a})^{\beta+n_{1}}
  (c,\nabla)^{\gamma-n_{3}-n_{1}-2(k_{3}+\tilde{k_{3}})}\tilde{\Theta}[\beta,\gamma,n_{1},p_{3},k_{3},\tilde{p_{3}},\tilde{k_{3}},c^{2},H_{3}]h^{(s)}(a^{\mu}) \nonumber\\\label{s3.19}
\end{eqnarray}
where
\begin{eqnarray}
  && \tilde{\Theta}[\gamma,\alpha,n_{2},p_{1},k_{1},\tilde{p_{1}},\tilde{k_{1}},a^{2},H_{1}]=
  \Theta[\gamma,\alpha,m_{1}=\tilde{p_{1}}-2\tilde{k_{1}}-n_{2},p_{1},k_{1},\tilde{p_{1}},\tilde{k_{1}},a^{2},H_{1}]\qquad\qquad \label{s3.20} \\
  && \tilde{\Theta}[\alpha,\beta,n_{3},p_{2},k_{2},\tilde{p_{2}},\tilde{k_{2}},b^{2},H_{2}]=
  \Theta[\alpha,\beta, m_{2}=\tilde{p_{2}}-2\tilde{k_{2}}-n_{3},p_{2},k_{2},\tilde{p_{2}},\tilde{k_{2}},b^{2},H_{2}] \label{s3.21} \\
  && \tilde{\Theta}[\beta,\gamma,n_{1}, p_{3},k_{3},\tilde{p_{3}},\tilde{k_{3}},c^{2},H_{3}]
  =\Theta[\beta,\gamma,m_{3}=\tilde{p_{3}}-2\tilde{k_{3}}-n_{1},\tilde{p_{3}}, p_{3},k_{3},\tilde{k_{3}},c^{2},H_{3}]\label{s3.22}
\end{eqnarray}
Taking into account that $\Theta[\dots, a^{2},H_{1}]\sim (a^{2})^{k_{1}+\tilde{k_{1}}}$ we see that our star products in (\ref{s3.19}) contract correctly all auxiliary vectors $a^{\mu}, b^{\nu}, c^{\lambda}$.

Then to understand better the structure of the derivativesof interaction we can take into account constraints (\ref{s3.16})-(\ref{s3.18}) and rearrange the summations coming from (\ref{s3.19}) in the following way
\begin{eqnarray}
 &&\sum_{n_{3}\geq 0}\sum_{n_{2}\geq 0}\sum_{n_{1}\geq 0} (-1)^{n_{1}+n_{2}+n_{3}}=\sum_{N\geq 0}(-1)^{N}\sum_{{n_{1}, n_{2}, n_{3}\atop \sum n_{i}=N}} ,\label{s3.23}\\
&&\sum_{{\{p_{i},k_{i},\tilde{p_{i}},\tilde{k_{i}}\}_{i=1,2,3}\atop p_{i}+\tilde{p_{i}}=n_{i}+n_{i+1}+2(k_{i}+\tilde{k_{i}})}}=\sum_{K\geq 0}\sum_{{{\{P_{i},K_{i}\}_{i=1,2,3}\atop P_{i}=n_{i}+n_{i+1}+2K_{i}}\atop \sum K_{i}=K}} \sum_{{\{p_{i},k_{i},\tilde{p_{i}},\tilde{k_{i}}\}_{i=1,2,3}\atop p_{i}+\tilde{p_{i}}=P_{i}; k_{i}+\tilde{k_{i}}=K_{i}}}\label{s3.24}
\end{eqnarray}
where in last equation $\{n_{i}\}=n_{1},n_{2},n_{3}$ with cyclic property $n_{4}=n_{1}$

After that we should introduce instead of $\alpha, \beta, \gamma$ new summation variables
\begin{eqnarray}
  \tilde{\alpha} &=&\alpha-n_{1}-n_{2}-2K_{1}=\alpha-P_{1},\label{s3.25}\\
  \tilde{\beta}&=&\beta-n_{2}-n_{3}-2K_{2}=\beta-P_{2},\label{s3.26}\\
  \tilde{\gamma}&=&\gamma-n_{3}-n_{1}-2K_{3}=\gamma-P_{3}.\label{s3.27}
\end{eqnarray}
with corresponding summation limits and constraints
 \begin{eqnarray}
 &&0\leq \tilde{\alpha}, \tilde{\beta}, \tilde{\gamma} \leq s-2(N+K) ,\label{s3.28}\\
 && \tilde{\alpha}+\tilde{\beta}+\tilde{\gamma}=s-2(N+K) ,\label{s3.29}\\
 && N=\sum_{i}n_{i}; \quad K=\sum_{i}K_{i}=\sum_{i}(k_{i}+\tilde{k_{i}}) .\label{s3.30}
 \end{eqnarray}
These transformations lead to the following formula:
\begin{eqnarray}
  &&\mathcal{L}_{I}^{main}=\int du e^{(d+2s-4)u} d^{d+1}x\sqrt{g}\sum_{N\geq 0}\sum_{K\geq 0}\frac{(-1)^{N}s!}{(s-2(N+K))!}\sum_{{\tilde{\alpha},\tilde{\beta},\tilde{\gamma}\atop \tilde{\alpha}+\tilde{\beta}+\tilde{\gamma}= s-2(N+K)}}
\binom{s-2(N+K)}{ \tilde{\alpha},\tilde{\beta},\tilde{\gamma}}\nonumber\\
&&\sum_{{\{n_{i}\}_{i=1,2,3}\atop \sum n_{i}=N}}\sum_{{{\{P_{i},K_{i}\}_{i=1,2,3}\atop P_{i}=n_{i}+n_{i+1}+2K_{i}}\atop \sum K_{i}=K}} \sum_{{\{p_{i},k_{i},\tilde{p_{i}},\tilde{k_{i}}\}_{i=1,2,3}\atop p_{i}+\tilde{p_{i}}=P_{i}; k_{i}+\tilde{k_{i}}=K_{i}}}\frac{*^{\tilde{\gamma}+\tilde{\alpha}+N+2(K_{3}+K_{1})} _{a^{\mu}}*^{\tilde{\alpha}+\tilde{\beta}+N+2(K_{1}+K_{2})} _{b^{\mu}}*^{\tilde{\beta}+\tilde{\gamma}+N+2(K_{2}+K_{3})} _{c^{\mu}} }{\binom{\tilde{\gamma}+\tilde{\alpha}+N+2(K_{3}+K_{1})+n_{1}}{n_{1}}\binom{\tilde{\alpha}+\tilde{\beta}+N+2(K_{1}+K_{2})+n_{2}}{n_{2}}
\binom{\tilde{\beta}+\tilde{\gamma}+N+2(K_{2}+K_{3})+n_{3}}{n_{3}}}
 \nonumber\\
 && (a,\partial_{b})^{\tilde{\gamma}+N+2K_{3}}
  (a,\nabla)^{\tilde{\alpha}} \,\Xi^{2K_{1}}[\tilde{\gamma},\tilde{\alpha},n_{2},p_{1},k_{1},\tilde{p_{1}},\tilde{k_{1}},a^{2},H_{1}]h^{(s)}(b^{\mu}) \nonumber\\
&&(b,\partial_{c})^{\tilde{\alpha}+N+2K_{1}}
  (b,\nabla)^{\tilde{\beta}}
  \,\Xi^{2K_{2}}[\tilde{\alpha},\tilde{\beta},n_{3},p_{2},k_{2},\tilde{p_{2}},\tilde{k_{2}},b^{2},H_{2}]h^{(s)}(c^{\mu}) \nonumber\\
  &&(c,\partial_{a})^{\tilde{\beta}+N+2K_{2}}
  (c,\nabla)^{\tilde{\gamma}}
  \,\Xi^{2K_{3}}[\tilde{\beta},\tilde{\gamma},n_{1},p_{3},k_{3},\tilde{p_{3}},\tilde{k_{3}},c^{2},H_{3}]h^{(s)}(a^{\mu}) \nonumber\\\label{s3.31}
\end{eqnarray}
where
\begin{eqnarray}
  && \tilde{\Theta}[\gamma,\alpha,n_{2},p_{1},k_{1},\tilde{p_{1}},\tilde{k_{1}},a^{2},H_{1}]=
  \frac{\gamma!}{\tilde{\alpha}!}\,\Xi^{2K_{1}}[\tilde{\gamma},\tilde{\alpha},n_{2},P_{3},p_{1},k_{1},\tilde{p_{1}},\tilde{k_{1}},a^{2},H_{1}] \qquad\qquad \label{s3.32} \\
  && \tilde{\Theta}[\alpha,\beta,n_{3},p_{2},k_{2},\tilde{p_{2}},\tilde{k_{2}},b^{2},H_{2}]=
  \frac{\alpha!}{\tilde{\beta}!}\,\Xi^{2K_{2}}[\tilde{\alpha},\tilde{\beta},n_{3},P_{1},p_{2},k_{2},\tilde{p_{2}},\tilde{k_{2}},b^{2},H_{2}]\label{s3.33} \\
  && \tilde{\Theta}[\beta,\gamma,n_{1}, p_{3},k_{3},\tilde{p_{3}},\tilde{k_{3}},c^{2},H_{3}]
  = \frac{\beta!}{\tilde{\gamma}!}\,\Xi^{2K_{3}}[\tilde{\beta},\tilde{\gamma},n_{1},P_{2},p_{3},k_{3},\tilde{p_{3}},\tilde{k_{3}},c^{2},H_{3}]\label{s3.34}
\end{eqnarray}
and
\begin{eqnarray}
  &&\Xi^{2K_{1}}[\tilde{\gamma},\tilde{\alpha},n_{2},P_{3},p_{1},k_{1},\tilde{p_{1}},\tilde{k_{1}},a^{2},H_{1}]\nonumber\\
  &&\quad=\frac{(\tilde{\alpha}+\tilde{p_{1}})!(a^2)^{k_{1}}}{(\tilde{\gamma}+P_{3}-\tilde{p_{1}}+2\tilde{k_{1}}+n_{2})!}
  \binom{\tilde{p_{1}}-2\tilde{k_{1}}}{n_{2}}\xi^{p_{1}+1}_{k_{1}}(\tilde{\alpha}+\tilde{p_{1}})
W^{\tilde{k_{1}}}(a^{2},H_{1})
   \,,\label{s3.35} \\
  && \Xi^{2K_{2}}[\tilde{\alpha},\tilde{\beta},n_{3},P_{1},p_{2},k_{2},\tilde{p_{2}},\tilde{k_{2}},b^{2},H_{2}]\nonumber\\
  &&\quad=\frac{(\tilde{\beta}+\tilde{p_{2}})!(a^2)^{k_{2}}}{(\tilde{\alpha}+P_{1}-\tilde{p_{2}}+2\tilde{k_{2}}+n_{3})!}
  \binom{\tilde{p_{2}}-2\tilde{k_{2}}}{n_{3}}\xi^{p_{2}+1}_{k_{2}}(\tilde{\beta}+\tilde{p_{2}})
W^{\tilde{k_{2}}}(b^{2},H_{2})\,,\label{s3.36} \\
  && \Xi^{2K_{3}}[\tilde{\beta},\tilde{\gamma},n_{1},P_{2},p_{3},k_{3},\tilde{p_{3}},\tilde{k_{3}},c^{2},H_{3}]\nonumber\\
  &&\quad=\frac{(\tilde{\gamma}+\tilde{p_{3}})!(a^2)^{k_{3}}}{(\tilde{\beta}+P_{2}-\tilde{p_{3}}+2\tilde{k_{3}}+n_{1})!}
  \binom{\tilde{p_{3}}-2\tilde{k_{3}}}{n_{1}}\xi^{p_{3}+1}_{k_{3}}(\tilde{\gamma}+\tilde{p_{3}})
W^{\tilde{k_{3}}}(c^{2},H_{3})\,.\label{s3.37}
\end{eqnarray}

Finalizing our consideration we can write direct $(a^{2}),(b)^{2},(c)^{2}$ expansion of corresponding $\Xi^{2K_{i}}$ terms using (\ref{s2.56}) and (\ref{s2.70})
\begin{eqnarray}
  && (a^2)^{k_{1}}W^{\tilde{k_{1}}}(a^{2},H_{1})h^{(s)}(b^{\mu})=
  \sum^{\tilde{k_{1}}}_{t_{1}=0}(-1)^{t_{1}}\sum^{\tilde{k_{1}}-t_{1}}_{r_{1}=1}
  \eta^{t_{1}}_{\tilde{k_{1}}}(\tilde{p_{1}}-\tilde{k_{1}})A^{\tilde{k_{1}}-t_{1}}_{r_{1}} [s]_{r_{1}} (a^{2})^{K_{1}-r_{1}}(a,b)^{r_{1}}\Phi_{r_{1}}(a,b)\qquad\qquad \label{s3.38} \\
 && (b^2)^{k_{2}}W^{\tilde{k_{2}}}(b^{2},H_{2})h^{(s)}(c^{\mu})=
  \sum^{\tilde{k_{2}}}_{t_{2}=0}(-1)^{t_{2}}\sum^{\tilde{k_{2}}-t_{2}}_{r_{2}=1}
  \eta^{t_{2}}_{\tilde{k_{2}}}(\tilde{p_{2}}-\tilde{k_{2}})A^{\tilde{k_{2}}-t_{2}}_{r_{2}}[s]_{r_{2}} (b^{2})^{K_{2}-r_{2}}(b,c)^{r_{2}}\Phi_{r_{2}}(b,c)\qquad\qquad \label{s3.39} \\
  && (c^2)^{k_{3}}W^{\tilde{k_{3}}}(a^{3},H_{3})h^{(s)}(c^{\mu})=
  \sum^{\tilde{k_{3}}}_{t_{3}=0}(-1)^{t_{3}}\sum^{\tilde{k_{3}}-t_{3}}_{r_{3}=1}
  \eta^{t_{3}}_{\tilde{k_{3}}}(\tilde{p_{3}}-\tilde{k_{3}})A^{\tilde{k_{3}}-t_{3}}_{r_{3}}[s]_{r_{3}} (c^{2})^{K_{3}-r_{3}}(c,a)^{r_{3}}\Phi_{r_{3}}(c,a)\qquad\qquad \label{s3.40}
\end{eqnarray}
So we see that $\Xi^{2K_{i}}$ in (\ref{s3.32})-(\ref{s3.34}) really behave like  $a^{2K_{1}}, b^{2K_{2}}, c^{2K_{3}}$ as they should for correct contractions of indices.

\section{Conclusion}
\quad We have constructed all $AdS$ corrections including trace and divergence terms to the main term of the cubic self-interaction  by a slightly  modified method of radial pullback (reduction) proposed in \cite{Manvelyan:2012ww} where all quantum fields are carried by a real  AdS space and corresponding interaction terms expressed through the covariant $AdS$ derivatives.
For given spin s and $\Delta_{min} = s$  we derived all curvature  correction terms (\ref{s3.31}) in the form of series of terms with numbers $s-2(N+K)$ of derivatives, where  $0 \leq N+K \leq \frac{s}{2}$. The latter is the number of seized pair of derivatives replaced by corresponding power of $1/L^{2}$ and $K$ is the sum of power of $a^{2}, b^{2}, c^{2}$ terms connected with trace and divergent correction terms produced from the main term of interaction after pullback. Correction terms  appear with coefficients that are polynomials in the dimension $d+1$ and spin number $s$  with rational coefficients. Now we can expect that the same method can be used for the derivation of the $AdS$ corrections to traces and deDonder terms connected with the main term by Noether's  procedure derived for the flat case in \cite{MMR1} and \cite{MMR2}.

\section*{Acknowledgements}
 \quad R.M. is indebted to Stefan Theisen  for valuable discussions and help during the visit in AEI. R.M. would like to thank Karapet Mkrtchyan and Ruben Mkrtchyan for discussion and comments. This work  were partially supported by the grant of the Science Committee of the Ministry of Science and Education of the Republic of Armenia under contract 18T-1C229.

\section*{Appendix A: Main Term of Cubic Interaction in Flat Space}
\setcounter{equation}{0}
\renewcommand{\theequation}{A.\arabic{equation}}
\quad In this Appendix we repeat the general formula for a covariant cubic interaction of higher spin gauge fields in a flat background as presented in \cite{MMR1} and \cite{MMR2}.
The main result of \cite{MMR1, MMR2} is the following. The gauge invariance fixes in a unique way the cubic interaction if the main cyclic ansatz term without divergences and traces is given. Accordingly in this article we consider only the main term of the cubic interaction postponing the proof for all other terms to a future publication, and understanding intuitively that gauge invariance is going to regulate in a correct fashion the radial reduction for all other terms presented in \cite{MMR1, MMR2} and classified in corresponding tables there.

In \cite{MMR1, MMR2} we considered three potentials $h^{(s_{1})}(X_{1};a^{A}), h^{(s_{2})}(X_{2};b^{A}), h^{(s_{3})}(X_{3};c^{A})$ of $d+2$ dimensional flat theory with ordered  spins $s_{i}$
\begin{equation}
s_{1} \geq s_{2}\geq s_{3} ,\label{A.1}
\end{equation}
and with the cyclic ansatz for the interaction
\begin{eqnarray}
&&\mathcal{L}_{I}^{main}(h^{(s_{1})}(X_{1},a^{A}),h^{(s_{2})}(X_{2},b^{A}),h^{(s_{3})}(X_{3}c^{A}))\nonumber\\
&&\hspace{2cm}=\sum_{n_{i}} C_{n_{1},n_{2},n_{3}}^{s_{1},s_{2},s_{3}} \int d^{d+2}X_{1}d^{d+2}X_{2}d^{d+2}X_{3} \delta (X_{3}-X_{1}) \delta(X_{2}-X_{1})\nonumber\\
&&\hspace{2cm}\times\tilde{T}(Q_{12},Q_{23},Q_{31}|n_{1},n_{2},n_{3})h^{(s_{1})}
(X_{1};a^{A})h^{(s_{2})}(X_{2};b^{B})h^{(s_{3})}(X_{3};c^{C}) ,\nonumber\\\label{A.2}
\end{eqnarray}
where
\begin{eqnarray}
&&\tilde{T}(Q_{12},Q_{23},Q_{31}|n_{1},n_{2},n_{3})\nonumber\\ &&\hspace{2cm}=(\partial_{a^{A}}\partial_{b_{A}})^{Q_{12}}(\partial_{b^{B}}\partial_{c_{B}})^{Q_{23}} (\partial_{c^{C}}\partial_{a_{C}})^{Q_{31}}(\partial_{a^{D}}\tilde{\nabla}_{2}^{D})^{n_{1}}(\partial_{b^{E}}\tilde{\nabla}^{E}_{3})^{n_{2}}( \partial_{c^{F}}\tilde{\nabla}^{F}_{1})^{n_{3}} ,\nonumber\\\label{A.3}
\end{eqnarray}
and the notation $"main"$ as a superscript means that it is an ansatz for terms without $Divh^{(s_{i}-1)}$ and $Trh^{(s_{i}-2)}$.
Denoting the number of derivatives by $\Delta$ we have
\begin{equation}
n_{1}+n_{2}+n_{3} = \Delta . \label{A.4}
\end{equation}
We shall later determine and then use the minimal possible $\Delta$. As balance equations we have
\begin{eqnarray}
n_{1}+Q_{12}+Q_{31} = s_{1} , \nonumber\\
n_{2}+Q_{23}+Q_{12} = s_{2} ,\nonumber\\
n_{3} + Q_{31} + Q_{23} = s_{3} .\label{A.5}
\end{eqnarray}
These equations are solved by
\begin{eqnarray}
Q_{12} = n_{3}-\nu_{3} ,\nonumber\\
Q_{23} = n_{1} - \nu_{1} , \nonumber\\
Q_{31} = n_{2} - \nu_{2} .\label{A.6}
\end{eqnarray}
Since the l.h.s. cannot be negative, we have
\begin{equation}
n_{i} \geq  \nu_{i} .\nonumber
\end{equation}
The $\nu_{i}$ are determined to be
\begin{equation}
\nu_{i} = 1/2 (\Delta +s_{i} -s_{j} -s_{k}), \quad i,j,k \quad  \textnormal{are all different.}\label{A.7}
\end{equation}
It follows that the minimally possible $\Delta$ is expressed by Metsaev's  \cite{Metsaev} (using the ordering of the $s_{i}$).
\begin{equation}
\Delta_{min} = \max{[s_{i} +s_{j} -s_{k}]} = s_{1}+ s_{2} -s_{3} .\label{A.8}
\end{equation}
Another result of \cite{MMR1, MMR2} is the trinomial expression for the coefficients in (\ref{A.2}) fixed by Noether's procedure. Taking into account (\ref{A.5})-(\ref{A.8}) we can write it in the following elegant form
\begin{eqnarray}
  C_{n_{1},n_{2},n_{3}}^{s_{1},s_{2},s_{3}}&=& C_{Q_{12},Q_{23},Q_{31}}^{s_{1},s_{2},s_{3}} = const\quad {s_{min} \choose Q_{12},Q_{23},Q_{31}} .\label{A.9}
\end{eqnarray}

\section*{Appendix B:Computation of $\xi^{p+1}_{k}$ for  $p=1,\dots , 4$ }
\setcounter{equation}{0}
\renewcommand{\theequation}{B.\arabic{equation}}
We start the computation of the following expression using iterative approach for different values of $p$
\begin{equation}
 V^{p+1}(i_{p+1})= \sum^{i_{p+1}}_{i_{p=0}}\phi_{i_{p}} V^{p}(i_{p}),
\end{equation}
where
\begin{equation}
\phi_{i_{k}}=a^{u}[2(i_{k}+k)-s]+[a^{2}-(a^{u})^{2}]\partial_{a^{u}}.
\end{equation}
After the computation, we can expand the result using the following formula
\begin{equation}
V^{p+1}(i_{p+1})=\sum_{k=0}^{[\frac{p}{2}]} \xi^{p+1}_{k}(i_{p+1}) (a^2)^{k} (a^{u}) ^{p-2k}.
\end{equation}
 and obtain $\xi^{p+1}_{k}$ coefficients.
\begin{flalign}
V^{2}(i_{2}) = \sum_{i_1=0}^{i_2} \phi_{i_1} |0> =  \left(1+i_2\right) \left(2-s+i_2\right) a^u|0> &&
\end{flalign}

\begin{flalign}
V^{3}(i_{3})=\sum_{i_1=0}^{i_2} \phi_{i_2}V^{2}(i_{2}) =\frac{1}{6} a^2 \left(1+i_3\right) \left(2+i_3\right) \left(6-3 s+2
   i_3\right)|0> &&
\end{flalign}
\begin{equation*}
+\frac{1}{2} \left(1+i_3\right) \left(2+i_3\right) \left(2-s+i_3\right)
   \left(3-s+i_3\right) \left(a^u\right)^2|0>
\end{equation*}

\begin{flalign}
V^{4}(i_{4}) = \sum_{i_3=0}^{i_4}  \phi_{i_3}V^{3}(i_{3}) =\frac{1}{6} a^2 \left(1+i_4\right) \left(2+i_4\right) \left(3+i_4\right)
   \left(4-s+i_4\right) \left(6-3 s+2 i_4\right) a^u|0> &&
\end{flalign}
\begin{equation*}
+\frac{1}{6} \left(1+i_4\right)
   \left(2+i_4\right) \left(3+i_4\right) \left(2-s+i_4\right) \left(3-s+i_4\right)
   \left(4-s+i_4\right) \left(a^u\right)^3|0>
\end{equation*}

\begin{flalign}
V^{5}(i_{5}) = \sum_{i_4=0}^{i_5}\phi_{i_4}V^{4}(i_{4}) =\frac{1}{360} a^4 \left(1+i_5\right) \left(2+i_5\right) \left(3+i_5\right)
   \left(4+i_5\right) &&
\end{flalign}
\begin{equation*}
 \left(360-270 s+45 s^2+172 i_5-60 s i_5+20
   i_5^2\right)|0>
\end{equation*}
\begin{equation*}
+\frac{1}{12} a^2 \left(1+i_5\right) \left(2+i_5\right) \left(3+i_5\right)
   \left(4+i_5\right) \left(4-s+i_5\right) \left(5-s+i_5\right) \left(6-3 s+2 i_5\right)
   \left(a^u\right)^2|0>
\end{equation*}
\begin{equation*}
+\frac{1}{24} \left(1+i_5\right) \left(2+i_5\right)
   \left(3+i_5\right) \left(4+i_5\right) \left(2-s+i_5\right) \left(3-s+i_5\right)
   \left(4-s+i_5\right) \left(5-s+i_5\right) \left(a^u\right)^4|0>
\end{equation*}
After investigating the structures of $\xi^{p+1}_{k}$ coefficients we notice that they all have the following general factor
\begin{eqnarray}
\frac{1}{(p-2k)!}(i+1)_{p}(2k+2+i-s)_{p-2k}
\end{eqnarray}
Using this information we can write the following ansatz for  $\xi^{p+1}_{k}$
\begin{eqnarray}
\xi^{p+1}_{k}(i) = \frac{1}{(p-2k)!}(i+1)_{p}(2k+2+i-s)_{p-2k}P_{k}(i)
\end{eqnarray}
Where $P_{k}(i)$ is $p$ independent polynomial.

\section*{Appendix C:The structure of the polynomial coefficients and the iterative approach of finding solutions}
\setcounter{equation}{0}
\renewcommand{\theequation}{C.\arabic{equation}}
To gain more information about the structure of polynomial coefficients we compute the following expression for initial values of $\tilde{k}=1,2,3,4 \dots$
\begin{eqnarray}
  &&W^{\tilde{k}+1}(a^{2},H,i_{\tilde{k}+2})=\sum^{i_{\tilde{k}+2}}_{i_{\tilde{k}+1}=1}\psi_{i_{\tilde{k}+1}-\tilde{k}}W^{\tilde{k}}
  (a^{2},H,i_{\tilde{k}+1})
\end{eqnarray}
where
\begin{equation}
\psi_{i} = i H + [i]_2\, a^2
\end{equation}
and factorize the resulting polynomial of variables $H$ and $a^2$. Then using the following expansion
\begin{eqnarray}
  && W^{\tilde{k}}(a^{2},H,i_{\tilde{k}+1})=\sum^{\tilde{k}}_{m=0}\eta^{m}_{\tilde{k}}(i_{\tilde{k}+1})(a^{2})^{m}H^{\tilde{k}-m}
\end{eqnarray}
 we can get the $\eta^{m}_{\tilde{k}}(i_{\tilde{k}+1})$ coefficients

\begin{flalign}\label{C.4}
W^{1}(a^{2},H,i_{2}) = \sum_{i_1=1}^{i_2}\psi_{i_1}\Phi_0(b)= \frac{1}{2} H i_2 \left(1+i_2\right) \Phi _0+\frac{1}{3} a^2 \left(-1+i_2\right) i_2 \left(1+i_2\right) \Phi _0 &&
\end{flalign}

\begin{flalign}
W^{2}(a^{2},H,i_{3}) = \sum_{i_2=1}^{i_3}\psi_{i_2-1}W^{1}(a^{2},H,i_{2})=\frac{1}{8} H^2 \left(-1+i_3\right) i_3 \left(1+i_3\right) \left(2+i_3\right) \Phi _0 &&
\end{flalign}
\begin{equation*}
+\frac{1}{12} a^2 H \left(-1+i_3\right) i_3 \left(1+i_3\right) \left(2+i_3\right)
   \left(-3+2 i_3\right) \Phi _0
\end{equation*}
\begin{equation*}
+\frac{1}{90} a^4 \left(-2+i_3\right) \left(-1+i_3\right) i_3 \left(1+i_3\right) \left(2+i_3\right) \left(-3+5 i_3\right) \Phi _0
\end{equation*}

\begin{flalign}
W^{3}(a^{2},H,i_{4}) = \sum_{i_3=1}^{i_4}\psi_{i_3-2}W^{2}(a^{2},H,i_{3}) = &&
\end{flalign}
\begin{equation*}
 \frac{1}{48} H^3 \left(-2+i_4\right) \left(-1+i_4\right) i_4 \left(1+i_4\right) \left(2+i_4\right) \left(3+i_4\right) \Phi _0
\end{equation*}
\begin{equation*}
+\frac{1}{24} a^2 H^2 \left(-2+i_4\right){}^2
   \left(-1+i_4\right) i_4 \left(1+i_4\right) \left(2+i_4\right) \left(3+i_4\right) \Phi _0
\end{equation*}
\begin{equation*}
+\frac{1}{180} a^4 H \left(-2+i_4\right) \left(-1+i_4\right){}^2 i_4
   \left(1+i_4\right) \left(2+i_4\right) \left(3+i_4\right) \left(-13+5 i_4\right) \Phi _0
\end{equation*}
\begin{equation*}
+\frac{a^6 \left(-3+i_4\right) \left(-2+i_4\right) \left(-1+i_4\right) i_4
   \left(1+i_4\right) \left(2+i_4\right) \left(3+i_4\right) \left(-2-63 i_4+35 i_4^2\right) \Phi _0}{5670}
\end{equation*}

\begin{flalign}\label{C.7}
W^{4}(a^{2},H,i_{5}) = \sum_{i_4=1}^{i_5}\psi_{i_4-3}W^{3}(a^{2},H,i_{4})= &&
\end{flalign}
\begin{equation*}
\frac{1}{384} H^4 \left(-3+i_5\right) \left(-2+i_5\right) \left(-1+i_5\right) i_5 \left(1+i_5\right)
\end{equation*}
\begin{equation*}
\left(2+i_5\right) \left(3+i_5\right) \left(4+i_5\right) \Phi_0
\end{equation*}
\begin{equation*}
+\frac{1}{288} a^2 H^3 \left(-3+i_5\right) \left(-2+i_5\right) \left(-1+i_5\right) i_5 \left(1+i_5\right)
\end{equation*}
\begin{equation*}
\left(2+i_5\right) \left(3+i_5\right) \left(4+i_5\right)
   \left(-5+2 i_5\right) \Phi _0
\end{equation*}
\begin{equation*}
+\frac{1}{1440}a^4 H^2 \left(-3+i_5\right) \left(-2+i_5\right) \left(-1+i_5\right) i_5 \left(1+i_5\right) \left(2+i_5\right) \left(3+i_5\right)
   \left(4+i_5\right)
\end{equation*}
\begin{equation*}
\left(45-46 i_5+10 i_5^2\right) \Phi _0
\end{equation*}
\begin{equation*}
+\frac{1}{22680}a^6 H \left(-3+i_5\right) \left(-2+i_5\right) \left(-1+i_5\right) i_5 \left(1+i_5\right)
   \left(2+i_5\right) \left(3+i_5\right) \left(4+i_5\right)
\end{equation*}
\begin{equation*}
 \left(-195+731 i_5-441 i_5^2+70 i_5^3\right) \Phi _0
\end{equation*}
\begin{equation*}
+\frac{1}{340200} a^8 \left(-4+i_5\right) \left(-3+i_5\right)
   \left(-2+i_5\right) \left(-1+i_5\right) i_5 \left(1+i_5\right) \left(2+i_5\right) \left(3+i_5\right) \left(4+i_5\right)
\end{equation*}
\begin{equation*}
 \left(570+149 i_5-630 i_5^2+175 i_5^3\right)
   \Phi _0
\end{equation*}

Examining the $\eta^{m}_{\tilde{k}}(i)$ coefficients we can see that they have the following form
\begin{equation}
 \eta^{m}_{\tilde{k}}(i) = \frac{2^{m - \tilde{k}}3^{-m}}{(\tilde{k} - m)!m!} (i-\tilde{k}+1)_{2\tilde{k}}P_m(i,\tilde{k}), \quad P_0(i,\tilde{k})=1
\end{equation}

It can be shown that $P_k(i, p)$ polynomials satisfy the following recurrent relation
\begin{equation}
(i + p + 1)P_k(i,p+1) - (i - p - 1)P_k( i - 1, p + 1) =
\end{equation}
\begin{equation}
2 (p - k + 1)P_k(i, p) + 3k (i - p - 1)P_{k - 1}( i, p)
\end{equation}

The solutions of this recurrent equation can be calculated step by step from the (\ref{C.4})-(\ref{C.7})  for each $k$

\begin{flalign}
P_0(i,p)=1 &&
\end{flalign}

\begin{flalign}
P_1(i, p) =i-\left(\frac{p}{2}+\frac{1}{2}\right) &&
\end{flalign}

\begin{flalign}
P_2(i, p) =i^2-2 i \left(\frac{p}{2}+\frac{3}{10}\right)+\left(\frac{p^2}{4}+\frac{3
   p}{20}-\frac{1}{10}\right) &&
\end{flalign}

\begin{flalign}
P_3(i, p) =i^3-3 i^2 \left(\frac{p}{2}+\frac{1}{10}\right)+3 i
   \left(\frac{p^2}{4}-\frac{p}{20}-\frac{67}{210}\right)-\left(\frac{p^3}{8}-\frac{3
   p^2}{20}-\frac{173 p}{280}-\frac{12}{35}\right) &&
\end{flalign}

\begin{flalign}
P_4(i,p)=i^4-4 i^3 \left(\frac{p}{2}-\frac{1}{10}\right)+6 i^2
   \left(\frac{p^2}{4}-\frac{p}{4}-\frac{481}{1050}\right) &&
\end{flalign}
\begin{equation*}
-4 i \left(\frac{p^3}{8}-\frac{3 p^2}{10}-\frac{1031
   p}{1400}-\frac{38}{175}\right)+\frac{p^4}{16}-\frac{11 p^3}{40}-\frac{2011
   p^2}{2800}-\frac{89 p}{1400}+\frac{111}{350}
\end{equation*}

\begin{flalign}
P_5(i,p)=i^5-5 i^4 \left(\frac{p}{2}-\frac{3}{10}\right)+10 i^3
   \left(\frac{p^2}{4}-\frac{9 p}{20}-\frac{181}{350}\right) &&
\end{flalign}
\begin{equation*}
-10 i^2
   \left(\frac{p^3}{8}-\frac{9 p^2}{20}-\frac{147 p}{200}+\frac{9}{175}\right)+5 i
   \left(\frac{p^4}{16}-\frac{3 p^3}{8}-\frac{351 p^2}{560}+\frac{843
   p}{1400}+\frac{3131}{3850}\right)
\end{equation*}
\begin{equation*}
-\frac{p^5}{32}+\frac{9 p^4}{32}+\frac{421
   p^3}{1120}-\frac{1587 p^2}{1120}-\frac{15839 p}{6160}-\frac{12}{11}
\end{equation*}

From the solutions above we can see that the general ansatz for $P_k(i,p)$ has the following form
\begin{equation*}
P_k(i,p) = \sum_{n=0}^{k} i^{k-n}(-1)^{n}\binom{k}{n}B^{n}_{k}(p)
\end{equation*}

From the solutions above for different $P_k(i,p)$ it is possible to find the solutions for $B^{n}_{k}(p)$ as follows \footnote{In order to compute the $B^{n}_{k}(p)$ using this iterative approach one should compute and know the expressions of $P_m(i,p)$ for up to $m=2k$}

\begin{flalign}
B^1{}_k(p)=\frac{p}{2}-\frac{k}{5}+\frac{7}{10} &&
\end{flalign}

\begin{flalign}
B^2{}_k(p)=\frac{p^2}{4}+p
   \left(\frac{11}{20}-\frac{k}{5}\right)+\frac{k^2}{25}-\frac{44
   k}{105}+\frac{607}{1050} &&
\end{flalign}

\begin{flalign}
B^3{}_k(p)=\frac{p^3}{8}+p^2 \left(\frac{3}{10}-\frac{3 k}{20}\right)+p \left(\frac{3
   k^2}{50}-\frac{377 k}{700}+\frac{641}{1400}\right) &&
\end{flalign}

\begin{equation*}
-\frac{k^3}{125}+\frac{293
   k^2}{1750}-\frac{1313 k}{1750}+\frac{108}{175}
\end{equation*}

\begin{flalign}
B^4{}_k(p)=\frac{p^4}{16}+p^3 \left(\frac{1}{8}-\frac{k}{10}\right)+p^2 \left(\frac{3
   k^2}{50}-\frac{157 k}{350}+\frac{13}{112}\right) &&
\end{flalign}

\begin{equation*}
+p \left(-\frac{2 k^3}{125}+\frac{523
   k^2}{1750}-\frac{131 k}{125}+\frac{519}{1400}\right)+\frac{k^4}{625}-\frac{244
   k^3}{4375}+\frac{47728 k^2}{91875}-\frac{460722 k}{336875}+\frac{256957}{404250}
\end{equation*}

\begin{flalign}
B^5{}_k(p)=\frac{p^5}{32}+p^4 \left(\frac{1}{32}-\frac{k}{16}\right)+p^3
   \left(\frac{k^2}{20}-\frac{251 k}{840}-\frac{443}{3360}\right) &&
\end{flalign}

\begin{equation*}
+p^2\left(-\frac{k^3}{50}+\frac{23 k^2}{70}-\frac{2273 k}{2800}-\frac{267}{1120}\right)
\end{equation*}
\begin{equation*}
+p\left(\frac{k^4}{250}-\frac{223 k^3}{1750}+\frac{2969 k^2}{2940}-\frac{980587
   k}{539000}-\frac{97283}{646800}\right)
\end{equation*}
\begin{equation*}
-\frac{k^5}{3125}+\frac{439
   k^4}{26250}-\frac{24404 k^3}{91875}+\frac{23911 k^2}{16170}-\frac{52309518
   k}{21896875}-\frac{72612}{398125}
\end{equation*}

The final form of $\eta^{m}_{\tilde{k}}(i)$ coefficients will be
\begin{equation}
 \eta^{m}_{\tilde{k}}(i) = \frac{2^{m - \tilde{k}}3^{-m}}{(\tilde{k} - m)!m!} (i-\tilde{k}+1)_{2\tilde{k}}
  \sum_{n=0}^{m} i^{m-n}(-1)^{n}\binom{m}{n}B^{n}_{m}(\tilde{k})
\end{equation}
\section*{Appendix D: Mapping operator $(a,\partial_{b})^{p}$ to the product of $H$ and $a^{2}$}
\setcounter{equation}{0}
\renewcommand{\theequation}{D.\arabic{equation}}
This is the final exercise to get more freedom in writing of our cubic interaction after our "stripping" for $u$ components of auxiliary vectors. Investigating (\ref{s2.68}) and first operator in (\ref{s2.9}):
\begin{equation}\label{D.1}
  (a^{A}\partial_{b^{A}})^{Q}=\sum^{Q}_{q=0}
  \binom{Q}{q}(a^{u}\partial_{b^{u}})^{Q-q}(a,\partial_{b})^{q}
\end{equation}
we see that last thing to do is transform the power of $(a,\partial_{b})$
to $H$ and $a^{2}$ to write interaction without $(a,\partial_{b})$, hiding them then in $\Phi(a,b)$ . Note that starting from (\ref{s2.53}) operator $H$ effectively worked with only its second part:
\begin{equation}\label{D.2}
  H\equiv H(a,b)=-(a,b)(a,\partial_{b})
\end{equation}
due to the separation of all $b^{u}$ dependence to the left from $H$ dependent part in (\ref{s2.68}). Therefore we can write simple relation
\begin{equation}\label{D.3}
  (a,\partial_{b})^{p}=(-1)^{p}\left(\frac{1}{(a,b)}H\right)^{p}
\end{equation}
Then introducing ansatz for ordered power:
\begin{equation}\label{D.4}
  \left(\frac{1}{(a,b)}H\right)^{p}=\frac{1}{(a,b)^{p}}
  \sum^{p-1}_{k=0}\rho_{k}(p)H^{p-k}(a^{2})^{k}
\end{equation}
and taking into account commutator
\begin{equation}\label{D.5}
  [H, (a,b)^{-k}]=\frac{k a^{2}}{(a,b)^{k}}
\end{equation}
we arrive to the following simple  triangular recurrence relation for polynomials $\rho_{k}(p)$
\begin{equation}\label{D.6}
  \rho_{k}(p+1)=\rho_{k}(p)+p \rho_{k-1}(p)
\end{equation}
with boundary conditions:
\begin{equation}\label{D.7}
  \rho_{0}(p)=1, \quad\quad \rho_{p-1}(p)=(p-1)!
\end{equation}
Recurrence relation (\ref{D.6}) we can easily solve using generation function. Introducing formal variable $z$ with $|z|<1$
\begin{equation}\label{D.8}
  \rho_{k}(z)=\sum^{\infty}_{p=0}z^{p}\rho_{k}(p)
\end{equation}
 we obtain recursive equation:
 \begin{equation}\label{D.9}
    \rho_{k}(z)=\frac{z^{2}}{1-z}\frac{d}{dz} \rho_{k-1}(z)
 \end{equation}
with the simple solution due to  boundary value
  $ \rho_{0}(z)=(1-z)^{-1}$ :
\begin{equation}\label{D.10}
 \rho_{k}(z)=\left[\frac{z^{2}}{1-z}\frac{d}{dz} \right]^{k}\frac{1}{1-z}
\end{equation}
In another way we can write the same solution of (\ref{D.6}) in the form of multiple sums:
\begin{eqnarray}
  && \rho_{k}(p)=\sum^{p-1}_{i_{k}=k}i_{k}\sum^{i_{k}-1}_{i_{k-1}=k-1}i_{k-1}
  \dots \sum^{i_{3}-1}_{i_{2}=2}i_{2}\sum^{i_{2}-1}_{i_{1}=1}i_{1} \label{D.11}
\end{eqnarray}

\end{document}